\documentclass[iop,apj]{emulateapj}

\usepackage{amsmath}
\usepackage{mathrsfs}
\usepackage{color}

\usepackage{times}
\usepackage{color}
\journalinfo{The Astrophysical Journal, \textit{in press}}
\slugcomment{ApJ,  in press}

\shorttitle{Observational Signatures of Coronal Loop Heating}
\shortauthors{Dahlburg et al.}

\begin{document}

\title{Observational Signatures of Coronal Loop Heating and Cooling\\
Driven by Footpoint Shuffling}

\author{R. B. Dahlburg$^{1}$, G.~Einaudi$^2$, B. D. Taylor$^1$, I. Ugarte-Urra$^3$,
H. P. Warren$^4$, A. F. Rappazzo$^5$, and M. Velli$^6$}
\affil{\vspace{.2em}
$^1$LCP\&FD, Naval Research Laboratory, Washington, DC 20375, USA; rdahlbur@lcp.nrl.navy.mil \\
$^2$Berkeley Research Associates, Inc., Beltsville, MD 20705, USA\\
$^3$College of Science, George Mason University, Fairfax, VA 22030, USA\\
$^4$Space Science Division, Naval Research Laboratory, Washington, DC 20375, USA\\
$^5$Advanced Heliophysics, Pasadena, CA 91106, USA\\
$^6$EPSS, UCLA, Los Angeles, CA 90095, USA}

\begin{abstract}
The evolution of a coronal loop is studied by means of numerical
simulations of the fully compressible three-dimensional
magnetohydrodynamic equations using the HYPERION code.  The footpoints
of the loop magnetic field are advected by random
motions.  As a consequence the magnetic field in the loop is
energized and develops turbulent nonlinear dynamics characterized by
the continuous formation and dissipation of field-aligned current sheets:
energy is deposited at small scales where heating occurs.  Dissipation
is non-uniformly distributed so that only a fraction of the
coronal mass and volume gets heated at any time. Temperature and
density are highly structured at scales which, in the solar corona, remain
observationally unresolved: the plasma of our simulated loop
is multi-thermal, where highly dynamical hotter and cooler plasma strands are
scattered throughout the loop at sub-observational scales.
Numerical simulations of coronal loops of 50000 km length and  axial magnetic field intensities ranging from 0.01 to 0.04 Tesla are
presented. To
connect these simulations to observations we use the computed number
densities and temperatures to synthesize the intensities expected in
emission lines typically observed with the Extreme ultraviolet Imaging Spectrometer
(EIS) on Hinode. These intensities are used to compute differential
emission measure distributions using the Monte Carlo Markov Chain code,
which are very similar to those derived from observations of solar active regions.
We conclude that coronal heating is found to be strongly intermittent in space and time,
with only small portions of the  coronal loop being heated:
in fact, at any given time, most of the corona is cooling down.
\end{abstract}


\section{Introduction}

The solar magnetic field has long been recognized as playing a key role in the
transport, storage and release of energy from the photosphere to the corona
\citep{1960MNRAS.120...89G, 1981ApJ...246..331S}.  \cite{1972ApJ...174..499P,
1994ISAA....1.....P} proposed that photospheric motions set the coronal
magnetic field in ``dynamic non-equilibrium'', that leads to the formation of
current sheets on fast ideal timescales \citep{2013ApJ...773L...2R} where
magnetic reconnection releases energy in small impulsive heating events termed
``nanoflares" \citep{1988ApJ...330..474P}.  This process has been shown to have
the characteristics of magnetically dominated MHD turbulence
\citep{1996ApJ...457L.113E, 1997ApJ...484L..83D, 2003PhPl...10.3584D,
  2007ApJ...657L..47R, 2008ApJ...677.1348R}, where the out-of-equilibrium
magnetic field generates a broad-band small velocity that creates small scales
distorting the magnetic islands and pushing field lines together
\citep{2011PhRvE..83f5401R}.  Similar dynamics are also displayed in cold plasma
\citep{1996ApJ...467..887H} and full MHD simulations \citep{1996JGR...10113445G,
  2012A&A...544L..20D}.

A first connection to observations has been provided by the statistics of these
bursty dissipative events, that have been shown to follow a power-law behavior
in total energy, peak dissipation and duration with indices not far from those
determined observationally in X-rays \citep{1998ApJ...497..957G,
1998ApJ...505..974D}.

But to constrain any model, advance our understanding of coronal heating and
correctly interpret observations it is crucial to study the thermodynamics of
such a system.  Simulations of entire active regions allow the investigation of
the geometric properties of radiative emission and thermodynamical quantities
\citep[e.g., temperature, mass flows and average volumetric heating rates,][]
{2002ApJ...572L.113G, 2011A&A...531A..97Z, 2013A&A...555A.123B}, but their
coarse resolution at scales below energy injection (about the granular
scale $\sim 10^3$\,km), necessary to include an entire active region, do
not allow the full development of nonlinear dynamics leading to the
formation of strong current sheets where energy is deposited.

Magnetic reconnection is not directly observable in the corona because it has become
increasingly clear that the effective heating and particle acceleration
occurs at scales of the order of the ion (proton) inertial
length $d_i$, which for an ion density $n_i \sim 10^8$~cm$^{-3}$ becomes $d_i = c/\omega_{pi} \sim 23$~m
(the proton plasma frequency is $\omega_{pi} = \sqrt{4\pi n_i e^2/m_i}$, $c$ the speed of light,
$e$ the electron charge, and $m_i$ the proton mass), \emph{well below the
resolution limits of present instrumentation} --- to
date the highest spatial resolution achieved for direct observations of the
corona is approximately $150$\,km by the Hi-C imager
\citep{2013Natur.493..501C}. Additionally for typical active region
temperatures $\sim 10^6$~K and magnetic field intensities $\sim 50$~G
the ion gyroradius is of same order of magnitude as $d_i$.

\emph{What can be observed directly is radiation}. By
analyzing the spectral properties of the observed radiation it is possible to
infer some of the physical properties of the plasma in the solar upper atmosphere,
such as the number density and temperature distribution along the line of
sight. Thus comparisons between observations and models must focus on the
analysis of the spectral properties of the plasma.

Here we analyze results from the HYPERION compressible MHD code. HYPERION is a
parallelized Fourier collocation finite difference code with Runge-Kutta time
discretization that solves the compressible MHD equations with parallel thermal
conduction and radiation included \citep{2010AIPC.1216...40D,
  2012A&A...544L..20D}.  HYPERION is able to produce temperatures and number
densities obtained in a framework where the ``heating function'' is due only to
the resistive and viscous dissipation induced in the corona by the footpoint shuffling.
Recent simulations \citep{2012A&A...544L..20D} have shown that temperature is highly
structured at scales below observational resolution in loops whose
magnetic field lines are shuffled at their footpoints by random
photospheric-mimicking motions: temperature peaks around
current sheets forming similarly shaped structures, approximately elongated in
the strong guide field direction, surrounded by cooler plasma.

In this paper we use our simulations of resolved loops to return
predictions for simulated ``observables'', such as the number density and
differential emission measure distribution, that can be compared with
observations. There has been considerable interest in the temperature
distribution observed in coronal loops
\citep[e.g.,][]{2003A&A...406.1089D,2007ApJ...656..577A,2002ApJ...578L.161S,2008ApJ...686L.131W,
  2012ApJ...759..141W}.  Many of these studies have found relatively narrow
emission measure distributions, and it has been unclear how these observations
could be reconciled with theory.

We simulate loops of $50,000$\,km length and axial magnetic fields of 100,
200, and 400\,G. The resulting temperatures and densities are used to synthesize
the emission line intensities that the Extreme Ultraviolet (EUV) Imaging
Spectrometer (EIS, \citealt{2007SoPh..243...19C}) would observe. These
intensities are input into the same analysis software used in many observational
studies. For these first calculations we find very good agreement between the
emission measure distributions derived from the simulations and the general
trends in the distributions derived from data. The distributions are relatively
narrow, peak at temperatures between $\log T = 6.0$ and 6.4, and show very
little emission at flare-like temperatures ($\log T\sim7$). The mean
temperature in the distribution, along with its width,
also rises with increasing field strength, consistent with observations.

\section{Formulation of the problem} \label{sec:fp}

In this section we describe our extension of the Parker coronal heating model
from RMHD to a formulation that includes many more significant physical processes.
We first describe our magnetohydrodynamic model in which physical augmentations, such
as thermal conduction and optically thin radiation, are contained.
Line-tied boundary conditions appropriate to the upper chromosphere are then given.
The velocity forcing function at the boundaries is also described.
The formulations for the elliptical gravity model, initial temperature and initial number density are 
also given.

\subsection{Governing equations}

We model the solar corona as a compressible, dissipative magnetofluid with nonlinear thermal conduction and optically thin radiation losses.
The governing equations, written here in dimensionless
form, are:
\begin{eqnarray}
{{\partial n}\over {\partial t}} &=& -\nabla\cdot (n {\bf v}),  \label{eq:eqn}      \\[.4em]
{{\partial n {\bf v}}\over{\partial t}} &=& -\nabla\cdot({n \bf v v})
   -{\beta}\nabla p + {\bf J}\times{\bf B}+
{1\over S_v}\nabla\cdot{\bf \zeta} \nonumber \\
&&+ \frac{1}{Fr^2}\   n \Gamma(z)\, {\bf\hat e}_z \label{eq:eqnv} \\[.4em]
{{\partial T}\over{\partial t}} &=& -{\bf v}\cdot\nabla T
 - (\gamma - 1) (\nabla\cdot {\bf v}) T
\nonumber \\
&& +\frac{1}{n} \Bigg\{ \frac{1}{ Pr\, S_v }
\bigg[{\bf B}\cdot\nabla
\bigg( \kappa_{\parallel}\ T^{5/2}\ {{\bf B}\cdot\nabla T\over B^2}\bigg)
\nonumber \\
&& +\kappa_{\perp} (n, \rho, T)\ \nabla\cdot
\bigg( {{\bf B}\times (\nabla T \times {\bf B})\over B^2}\bigg) \bigg]
\nonumber \\
&& + {(\gamma -1)\over\beta}  \bigg[
{ 1\over S_v} \zeta_{ij} {\partial v_i\over\partial x_j}
+{1\over S} (\nabla\times{\bf B})^2
\nonumber \\
&& -{1\over P_{rad} S_v} n^2\Lambda (T)
+ {\beta\over(\gamma - 1)} n C_N \bigg] \Bigg\}, \label{eq:eqT}\\[.4em]
{{\partial {\bf B}}\over{\partial t}} &=& \nabla\times{\bf v}\times{\bf B}
  - \frac{1}{S}\nabla\times \nabla\times {\bf B}, \label{eq:b}
  \end{eqnarray}
with the solenoidality condition $\nabla\cdot{\bf B} = 0$.
The system is closed by the equation of state
\begin{equation} \label{eq:eqp}
p = n T.
\end{equation}
The non-dimensional variables are defined in the following way:
$n ({\bf x}, t)$ is the number density,
${\bf v}({\bf x}, t) = (u, v, w)$ is the flow velocity,
$p({\bf x}, t)$ is the thermal pressure,
${\bf B}({\bf x}, t) = (B_x, B_y, B_z) $ is the magnetic induction field,
${\bf J} = \nabla\times{\bf B}$ is the electric current density,
$T({\bf x}, t)$ is the plasma temperature,
$\zeta_{ij}= \mu (\partial_j v_i + \partial_i v_j) -
\lambda \nabla\cdot {\bf v} \delta_{ij}$ is the viscous stress tensor,
$e_{ij}= (\partial_j v_i + \partial_i v_j)$ is the strain tensor,
and $\gamma$ is the adiabatic ratio.

To render the equations dimensionless we set characteristic values at the
walls of the computational box: a number density $n_*$,
vertical Alfv{\'e}n speed at the boundaries $V_{A*}$,
the orthogonal box width $L_*$,  and the temperature $T_*$.
Therefore time ($t$) is measured in units of the Alfv\'en time
($\tau_A=L_* /V_{A*}$ -- note that this is not the axial loop length transit time.).
The parallel thermal conductivity is given by $\kappa_\parallel$,
while the perpendicular thermal conduction is considered negligible and hence
$\kappa_{\perp}$ is set to zero.

The magnetic resistivity $\eta$, and shear viscosity $\mu$
are assumed to be constant and uniform, and Stokes relationship is assumed
so the bulk viscosity $\lambda = (2/3) \mu$.
In our previous paper \citep{2012A&A...544L..20D} the function
$\Lambda(T)$ that describes the temperature dependence of the radiation
was evaluated in the same way as \cite{1974SoPh...35..123H}.
Here we use instead the radiation function based on the CHIANTI atomic database
\citep{2012ApJ...744...99L}, normalized by its value at the base temperature
$T_* = 10000\, K$.
The Newton cooling term $C_N$ is described in section~\ref{sec:grav}.

The important dimensionless numbers are:
$S_v = n_* m_p  V_{A*} L_* / \mu \equiv$ viscous Lundquist number
($m_p = 1.673\times 10^{-27}$ kg is the proton mass),
$S = \mu_0 V_{A*} L_* / \eta \equiv$ Lundquist number
($\mu_0 = 1.256\times10^{-6}$ Henrys / meter is the magnetic permeability),
$\beta = \mu_0 p_* / B_*^2 \equiv$ pressure ratio at the wall,
$Pr = C_v \mu / \kappa_{\parallel} T_*^{5/2} \equiv$ Prandtl number, and
$P_{rad} $, the radiative Prandtl number
${\mu/ \tau_A^{2} n_*^2 \Lambda (T_*)} $. $C_v$ is
the specific heat  at constant volume.
The magnetohydrodynamic Froude number ($Fr$) is equal to $V_A/(g L_*)^{1/2}$,
where $g=274$~m~s$^{-2}$ is the solar surface gravity.

In what follows we assume normalizing quantities representative of the upper solar chromosphere:
$n_*=10^{17}\, $m$^{-3}$,
$T_*=10^{4}$ K,  and $L_* = 4 \times 10^{6}$ m.
$B_*$ is the only quantity that is varied in the three numerical simulations $(B_*=$0.01 Teslas;  0.02 Teslas and 0.04 Teslas; see Table~\ref{tab:table 1}).
We set $\ln \Lambda = 10$.  A loop length of $L_{z*}= 12.5 L_*$= 50000 km is used in all of the simulations.
The normalized time scale of the forcing, $t^*$, is set to represent
a five minute  convection time scale. The normalized
velocity $V_*$ is $10^3$~m~s$^{-1}$.  This velocity is expressed in dimensionless form as $\Xi=V_* / V_{A*}$.

\begin{figure}
\begin{centering}
\includegraphics[width=1.\columnwidth]{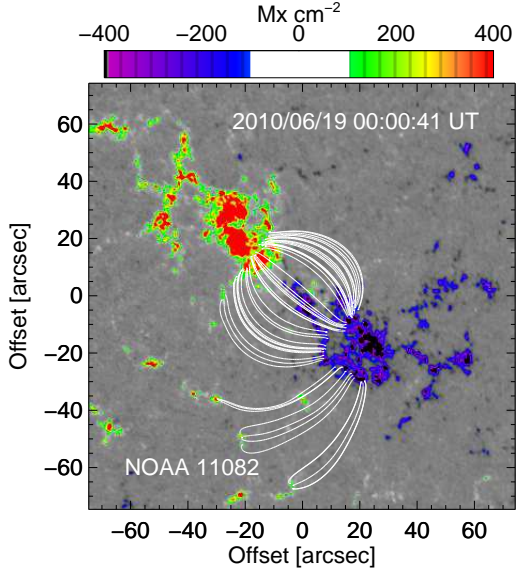}
\caption{An illustrative observation of a small solar active region. Here we
  show magnetic field lines with total lengths between 45 and 55\,Mm
  extrapolated from an HMI magnetogram. The magnetic field strengths are
  color coded.
\label{h1}}
\end{centering}
\end{figure}

\subsection{Boundary and initial conditions} \label{sec:inibc}

We solve the governing equations in a Cartesian domain of size
$L_x \times L_y \times L_z = 1 \times 1 \times L_z$,
where $L_z$ is the loop aspect ratio determined by the loop length
and the characteristic length ($0\le x,y \le1,\ -L_z/2 \le z \le L_z/2$).
The system has periodic boundary conditions in $x$ and $y$,  line-tied boundary
conditions at the top and bottom $z$-plates,
and it is threaded by a strong guide magnetic field $B_0 = 1$ in the $z$-direction.

As explained later in subsection \ref{subsec:nm}
we utilize the magnetic vector potential rather than the magnetic induction field.
In addition, our implementation of a staggered mesh in $z$ is explicated in subsection \ref{subsec:nm}
Using the normalizing quantities given above,
the dimensionless line-tied boundary conditions which are enforced at the top and bottom walls of the simulation take the following form:
\begin{equation}
n = 1,
\end{equation}
\begin{equation}
T = 1,
\end{equation}
\begin{equation}
nu = n{\partial\psi\over\partial y},
\end{equation}
\begin{equation}
nv = -n{\partial\psi\over\partial x},
\end{equation}
\begin{equation}
nw = 0,
\end{equation}
\begin{equation}
{\partial A_x\over\partial t} = v B_z,
\end{equation}
\begin{equation}
{\partial A_y\over\partial t} = -u B_z,
\end{equation}
and
\begin{equation}
B_z = 1.
\end{equation}
The velocity stream function ($\psi$) is described in section \ref{sec:vel}.
The magnetic field is expressed as
$\mathbf{B} = B_0 \mathbf{\hat e}_z + \mathbf{b}$
with $\mathbf{b} (x,y,z,t)=  \nabla \times \mathbf{A}$, where $\mathbf{A}$ is the vector
potential associated with the fluctuating magnetic field.
At the top and bottom $z$-plates $B_z$, $n$ and $T$
are kept constant at their initial values $B_0$, $n_0$ and $T_0$,
while the magnetic vector potential is convected by the resulting flows.

\subsubsection{Velocity forcing function} \label{sec:vel}
At the boundaries we employ a time-dependent
forcing function analogous to those used in previous studies
\citep{1996ApJ...467..887H, 1996ApJ...457L.113E, 1999PhPl....6.4146E},
i.e., at the top boundary $z=L_z/2$ we evolve a function
\begin{equation} \label{eq:fc1}
\phi_t (x, y, t) = f_1 \sin^2 \left(\frac{\pi t}{2 t^*}\right) + f_2  \sin^2 \left(\frac{\pi t}{2 t^*} + \frac{\pi}{2}\right),
\end{equation}
and at the bottom boundary $z=-L_z/2$ we evolve a similar function
\begin{equation} \label{eq:fc2}
\phi_b  (x, y, t) = f_3 \sin^2 \left(\frac{\pi t}{2 t^*}+ \frac{\pi}{4}\right) + f_4  \sin^2 \left(\frac{\pi t}{2 t^*} + \frac{3\pi}{4}\right),
\end{equation}
where
\begin{equation} \label{eq:fc3}
f_i (x,y) =  \sum_{m, p} \frac{a_{mp}^i\ \sin \left[2\pi (m x + p y + \chi_{mp}^i) \right]}
                 { \sqrt{m^2 + p^2} },
\end{equation}
in which all wave-numbers with $3\le \sqrt{m^2 + p^2} \le4$ are excited, so that
the typical length-scale of the eddies is $\sim 1/4$.
$a_{mp}^i$ and $\chi_{mp}^i$ are random numbers chosen such that
$0\le a_{mp}^i, \chi_{mp}^i \le1$.
Every $t^*$, the coefficients $a_{mp}^i$ and  $ \chi_{mp}^i$ are randomly
changed alternatively for eddies 1 through 4.

At each timestep a provisional wall velocity is computed from:
\begin{equation}
u_{prov}={\partial\phi\over\partial y}
\end{equation}
and
\begin{equation}
v_{prov}=-{\partial\phi\over\partial x}
\end{equation}
To ensure that the kinetic energy at the wall remains constant, we compute
\begin{equation}
K=\sum_{j=1}^{n_y}\sum_{i=1}^{n_x} \bigl[ u_{prov}^2(i,j) + v_{prov}^2(i,j) \bigr]
\end{equation}
separately at the top and bottom boundaries (these are denoted by $K_t$ and $K_b$).
To achieve the desired velocity we then have the following stream functions at the top and bottom boundaries:
\begin {equation}
\psi_t = {\Xi\over K_t } \phi_t
\end{equation}
and
\begin {equation}
\psi_b = {\Xi\over K_b} \phi_b,
\end{equation}
where $\Xi=V_* / V_{A*}$.
Based on these stream functions, the top boundary velocity is given by:
\begin{equation}
u_t={\partial\psi_t\over\partial y}~~{\rm and}~~v_t=-{\partial\psi_t\over\partial x},
\end{equation}
and the bottom boundary velocity is given by:
\begin{equation}
u_b={\partial\psi_b\over\partial y}~~{\rm and}~~v_b=-{\partial\psi_b\over\partial x}
\end{equation}

\begin{table*}
\begin{center}

\caption{\label{tab:table 1}  Dimensionless numbers based on solar values. }
\bigskip

\begin{tabular*}{\textwidth}{c @{\extracolsep{\fill}} ccccccc}
\hline \hline\noalign{\vspace{.5em}}
 Case & $B_0$ (Tesla) & $\beta$ & $S_v$ & $S$ & $Fr$  & $Pr$  & $P_{rad}$ \\[.6em]
\hline\noalign{\vspace{.5em}}
A&0.01&  $1.735\times 10^{-4}$&$2.088\times 10^9$& $2.694\times 10^9$ & $2.083\times 10^1$ & $2.533\times 10^{-2}$ & $7.339\times10^{-7}$ \\[.3em]
B&0.02&  $4.339\times 10^{-5}$&$4.176\times 10^9$& $5.389\times 10^9$ & $4.166\times 10^1$ & $2.533\times 10^{-2}$ & $2.935\times10^{-6}$ \\[.3em]
C&0.04&  $1.085\times 10^{-5}$&$8.352\times 10^9$& $1.078\times 10^{10}$ & $8.332\times 10^1$ & $2.533\times 10^{-2}$ & $1.174\times10^{-5}$ \\[.3em]
D&0.01&  $1.735\times 10^{-4}$&$2.088\times 10^9$& $2.694\times 10^{9}$ & $2.083\times 10^1$ & $2.533\times 10^{-2}$ & $7.339\times10^{-7}$ \\[.3em]
E&0.01&  $1.735\times 10^{-4}$&$2.088\times 10^9$& $2.694\times 10^9$ & $2.083\times 10^1$ & $2.533\times 10^{-2}$ & $7.339\times10^{-7}$ \\[.3em]
F&0.01&  $1.735\times 10^{-4}$&$2.088\times 10^9$& $2.694\times 10^9$ & $2.083\times 10^1$ & $2.533\times 10^{-2}$ & $7.339\times10^{-7}$ \\[.3em]
G&0.01&  $1.735\times 10^{-4}$&$2.088\times 10^9$& $2.694\times 10^9$ & $2.083\times 10^1$ & $2.533\times 10^{-2}$ & $7.339\times10^{-7}$ \\[.3em]
\end{tabular*}

\end{center}
\end{table*}

\begin{table*}
\begin{center}
\caption{\label{tab:table 2}  Numerical resolution and rescaled dimensionless numbers used in the numerical simulations. }
\bigskip

\begin{tabular*}{\textwidth}{c @{\extracolsep{\fill}} cccccc}
\hline \hline\noalign{\vspace{.5em}}
 CASE & Resolution ($n_x \times n_y \times n_z$) & $\tilde{R}$ & $\tilde{S}_v$ & $\tilde{S}$ & $\widetilde{Pr}$  & $\tilde{P}_{rad}$ \\[.6em]
\hline\noalign{\vspace{.5em}}
A&  $64\times64\times144$ &$50 $ & $3.448\times 10^4$ & $4.449\times 10^4$ &  $1.534\times 10^3$ & $4.444\times10^{-2}$ \\[.3em]
B&  $64\times64\times144$ &$50 $ & $6.896\times 10^4$ & $8.898\times 10^4 $ & $1.534\times 10^3$ & $1.778\times10^{-1}$ \\[.3em]
C&  $64\times64\times144$ &$50 $ & $1.379\times 10^5$ & $1.780\times 10^5 $ & $1.534\times 10^3$ & $7.111\times10^{-1}$ \\[.3em]
D&  $128\times128\times144$ &$100 $ & $6.896\times 10^4$ & $8.898\times 10^4 $ & $7.670\times 10^2$ & $2.222\times10^{-2}$ \\[.3em]
E&  $64\times64\times 288$ &$50 $ & $3.448\times 10^4$ & $4.449\times 10^4$ &  $1.534\times 10^3$ & $4.444\times10^{-2}$ \\[.3em]
F&  $64\times64\times 576$ &$50 $ & $3.448\times 10^4$ & $4.449\times 10^4$ &  $1.534\times 10^3$ & $4.444\times10^{-2}$ \\[.3em]
G&  $64\times64\times 1620$ &$50 $ & $3.448\times 10^4$ & $4.449\times 10^4$ &  $1.534\times 10^3$ & $4.444\times10^{-2}$ \\[.3em]
\end{tabular*}

\end{center}
\end{table*}

\subsubsection{Initial condition for dynamical variables}
As explained later in subsection \ref{subsec:nm}
we utilize the magnetic vector potential rather than the magnetic induction field.
The initial values for the momentum and magnetic vector potentlal are given by:
\begin{equation}
nu = 0,
\end{equation}
\begin{equation}
nv = 0,
\end{equation}
\begin{equation}
nw = 0,
\end{equation}
\begin{equation}
A_x = 0,
\end{equation}
\begin{equation}
 A_y = 0,
\end{equation}
and
\begin{equation}
A_z = 0.
\end{equation}

\subsubsection{Initial temperature, number density and gravity specification} \label{sec:grav}
Here we describe how we initialize the temperature and number density, as well as specify the gravity function.
The loop gravity is determined by an elliptical model, with
\begin{equation}
\Gamma(z)={b z \over a^2 (1-z^2/a^2)^{1/2}},
\end{equation}
where $a$ is the semi-major and $b$ is the semi-minor axis of the ellipse.
The elliptical model decouples the loop height from the loop length,
since $a$ and $b$ can be specified independently.
The footpoints of the loop are located where $d\Gamma/ dz=\pm 1$, i.e.,
the loop length is given by $L_z=2.0 \left[ a^4 / \left(a^2 + b^2 \right) \right]^{1/2}$.

We impose as initial condition a temperature profile ($T_i$) with the dimensionless temperature 1
at the boundaries and 100 in the center (this corresponds to dimensional values of $10^4$ K
and $10^6$ K.
Let $T_{apex} = 100$, then:
\begin{equation}
T_i (z)  = T_{apex} - {T_{apex}-1\over(0.5 L_z) ^ q}~~z^q.  \label{eq:tempinit}
\end{equation}
The parameter $q$ determines the steepness of the temperature profile at the boundaries, as well
as the flatness of the temperature profile in the center of the system.
We set $q=8$ to ensure a rapid increase of temperature away from the boundaries.
The number density can then be solved for in the usual manner, i.e.,
\begin{equation}
\frac{d}{dz} nT_i = n{dT_i\over dz} + T_i {dn\over dz} = {1\over \beta Fr^2} n \Gamma (z).
\end{equation}
Rearranging this equation, we have:
\begin{equation}
{1\over n} {dn\over dz} = \frac{d}{dz} {\rm ln} n =
- \frac{d}{dz} {\rm ln} T_i + {1\over \beta Fr^2} {1\over T_i} \Gamma (z).
\end{equation}
We solve this numerically with a shooting method.  Calculating the
number density in this way allows us to consider longer loops.  In this
paper we choose $a = b = 6.25 \sqrt{2}$, consistent with $L_z =
12.5$ (and a dimensional loop length of 50000 km).  Combined with our choices for $B_*$, this places our
loop within the range of what is typically observed in the solar corona.  To
illustrate this, in Figure~\ref{h1} we show active region NOAA 11082 with field lines
in the range 45000 -- 55000~km computed from a potential extrapolation of a Helioseismic
and Magnetic Imager \citep{2012SoPh..275..229S} magnetogram.

The term $C_N$ in equation \ref{eq:eqT} denotes a Newton cooling function which is enforced
close to the $z$ boundaries \citep{2001A&A...365..562D, bp11}.
In dimensionless form we use $C_N = {1\over\ \tau_N}~[T_i (z) - T(z)] e^{-(z+0.5L_z)/h_N}$ at the lower boundary and
$C_N = {1\over\ \tau_N}~[T_i (z)  - T(z)] e^{-(0.5L_z-z)/h_N}$ at the upper boundary.
Here $\tau_N$ is the Newton cooling time and $h$ is the Newton cooling height.
We use $\tau_N = 10$ and $h_N$ = 1/4.
In dimensional terms this corresponds to times between 0.145 s and 0.58 s for the various magnetic field cases (see Table 1),
and a height of 1000 km. 
The Newton cooling term is only effective over the first few points in $z$ at each boundary.
Note as well that the radiation function is exponentially decreased in the inverse manner near the boundary to account for the
increasing optical thickness of the upper chromosphere.
\section{Numerical considerations} \label{sec:nc}

\subsection{Numerical method} \label{subsec:nm}

With previous definitions equation~(\ref{eq:b}) and the magnetic field solenoidality condition
($\nabla\cdot{\bf B} = 0$) can be replaced by the
magnetic vector potential equation:
\begin{equation} \label{eq:eqbp}
 {\partial {\bf A}\over\partial t}={\bf v}\times ( B_0\, \mathbf{\hat e}_z + \nabla\times{\bf A} )
  - {1\over S}~ \nabla\times\nabla\times {\bf A}
\end{equation}

A staggered mesh is employed in the $z$-direction \citep{1987JCP..70...300T}.
The fields that are defined at the
$z$ boundaries are advanced in time on the standard mesh.  Other quantities of interest are defined and
advanced in time on the staggered mesh.
That is, on the standard mesh we evaluate $n, n u, ~n v, ~n w, ~A_x, ~A_y,~B_z$ and $T$.  Some
derived fields such as $\omega_x, ~\omega_y, ~\omega_z,  ~j_x,$ and $j_y$ are also defined on the standard
mesh.  On the staggered mesh we evaluate $A_z, ~B_x,~ B_y,$ and $j_z$.  Note that for plotting purposes we
interpolate these latter fields onto the standard mesh (at the boundaries an extrapolation is performed).

We solve numerically equations~(\ref{eq:eqn})-(\ref{eq:eqT}) and (\ref{eq:eqbp})
together with equation~(\ref{eq:eqp}).  When solving for the $z$ magnetic field we add
the DC magnetic field contribution to $(\nabla\times{\bf A})_z$.
Space is discretized in $x$ and
$y$ with a Fourier collocation scheme \citep{1989PhFlB...1.2153D} with isotropic
truncation dealiasing.
Spatial derivatives are calculated in Fourier space, and nonlinear product terms are
advanced in configuration space.
A second-order central difference technique on a uniform mesh is used for the
discretization in $z$ \citep{1986JFM...169...71D}.
Variables are advanced in time by a low-storage Runge-Kutta scheme.  Several
options are available: two-step second-order,
three-step third-order, four-step third-order and five-step fourth-order
\citep{carpenter1994fourth}.
Results presented in this paper use the last option, as it permits the largest time
step.
Thermal conduction is advanced with second-order Super TimeStepping
\citep{2012MNRAS.422.2102M}.

HYPERION, which previously used only MPI for parallel execution, was modified for hybrid parallelization using a combination of OpenMP and MPI.  The code retains its original MPI-only strategy of assigning groups of $x$--$y$ planes to each MPI rank by decomposing the three-dimensional simulation domain along the $z$ direction.  This keeps all of the data needed for FFTs in the periodic $x$ and $y$ directions local to each MPI rank.  Scalability of the original MPI-only code was limited, however, because the maximum number of MPI ranks that could be used in a given simulation could not exceed the number of $x$--$y$ planes in the domain.  In the hybrid code, OpenMP multithreading is used to exploit parallel work within the groups of $x$--$y$ planes assigned to each MPI rank, for example by computing one-dimensional FFTs in the $x$--$y$ planes in parallel.  This allows more CPU cores to be utilized than was possible with the MPI-only version and, for a fixed number of cores, reduces the overhead of MPI communication relative to the MPI-only code.

\subsection{Simulation rescaling}
The dimensionless numbers based on the physical parameters are given in Table 1.
Note that  physical Lundquist and Reynolds numbers are far too large for
present day computations.
Consider, for example, case A which has a characteristic flow velocity given by
$V_* = 1.0\times 10^3$ m~s$^{-1}$ and a characteristic
Alfv\'en velocity given by $V_{A*} = 6.896\times 10^5$~m~s$^{-1}$.
We have a physical Reynolds number equal to:
\begin{equation}
R={V_*\over V_{A*}} S_v = 3.028\times 10^6
\end{equation}
and a physical magnetic Reynolds number equal to:
\begin{equation}
R_m={V_*\over V_{A*}} S = 3.905\times 10^6
\end{equation}
(here ${V_*/ V_{A*}} = M_A = 1.45\times 10^{-3}$ can be thought of as an Alfv\'en Mach number).
Rather than use these numerically unresolvable Reynolds numbers
we present the results obtained running the code with smaller Reynolds
numbers that can be used with the currently achievable numerical
resolution. For example in case~A they are $\tilde{R} = 50$ and
$\tilde{R}_m  = (S/S_v)\tilde{R}= 64.51$,
with a horizontal resolution of $64^2$, i.e.,
for case~A we use
\begin{equation}
\tilde{S_v}  = {\tilde{R}\over M_A} = 3.448 \times 10^4
\end{equation}
and
\begin{equation}
\tilde{S} = {\tilde{R}_m\over M_A} = 4.449 \times 10^4.
\end{equation}
These somewhat conservative values of the Reynolds numbers are taken based on previous numerical simulations of
turbulent magnetofluids \citep{1989PhFlB...1.2153D}.
In order to keep the same relative efficiency of the radiative and conductive terms in the energy equation
as in the real corona, we have rescaled $Pr$ and $P_{rad}$ accordingly with the choice
of $\tilde{S_v}$, i.e., we set
\begin{equation}
\widetilde{Pr}~\tilde{S_v} = Pr~S_v
\end{equation}
and
\begin{equation}
\tilde{P}_{rad}~\tilde{S_v} = P_{rad}~S_v
\end{equation}
so that for case A:
$\widetilde{Pr} = 1.534 \times 10^3$
and
$\tilde{P}_{rad} = 4.444\times10^{-2}$.
This rescaling is motivated by the result found in the RMHD model
\citep{2008ApJ...677.1348R}
that turbulent dissipative processes  are independent of viscosity and
resistivity when an inertial range is well resolved.
The rescaled values are given in Table 2.

Numerical resolutions for all of the simulations are given in Table 2.
These resolutions are smaller than our previous RMHD simulations.
However, the present simulations integrate more complex governing
equations, evolving eight different field components (number density, temperature,
the magnetic vector potential field and the velocity field) compared to only
two scalar fields in RMHD.
In addition, the density stratification from the upper chromosphere to the corona
constrains us, at present, to compute with a very small time step due to the large
variation in the Alfv\'en speed along the loop.

\section{Results} \label{sec:res}

We here discuss the transmission, storage and release of
energy in the simulated coronal loop.
At $t=0$ the system starts out in a ground state, defined by
the constant initial axial magnetic field $B_0 \mathbf{\hat{e}}_z$,
zero magnetic field fluctuations $\mathbf{b}$, while the
initial number density and  temperature profiles are as described
in section~\ref{sec:grav}. The  velocity field vanishes everywhere
initially except at the top and bottom boundaries ($z = \pm L_z/2$) as described
in section~\ref{sec:vel}.
Radiation and thermal conduction are ramped up linearly until they
attain their full values at $t = t*$.

\subsection{Loop energization}

The random velocity fields at the top and bottom boundaries
twist the field lines in a disordered way (since the forcing velocity is
not symmetric), creating a magnetic field component $\mathbf{b}$
predominantly orthogonal to the DC magnetic field. Initially $\mathbf{b}$ evolves quasi-statically
thus growing linearly with time
\citep{2008ApJ...677.1348R, 2015arXiv150504370R}.
But as soon as the intensity of $\mathbf{b}$ grows beyond
a certain threshold, that depends on the loop parameters,
current sheets form on a fast ideal timescale, with their
width thinning down to the dissipative scale in about
an axial Alfv\'en transit time $\tau_A = L_z/V_A$
\citep{2013ApJ...773L...2R}.
Furthermore thinning current sheets have been recently
shown to be unstable to tearing modes with ``ideal'' (i.e.,
of the order of $\tau_A$) growth rates even for thicknesses
larger than Sweet-Parker \citep{2014ApJ...780L..19P}. Overall this implies that once
the field lines are twisted beyond a certain threshold, or equivalently
once the magnetic field intensity grows beyond a corresponding threshold,
the magnetic field is no longer in equilibrium and transitions on the
ideal timescale to a magnetically dominated MHD turbulence
regime, where magnetic fluctuations are stronger than velocity
fluctuations \citep{1996ApJ...457L.113E, 1997ApJ...484L..83D,
2011PhRvE..83f5401R}.

The work done by boundary motions on the magnetic field line
footpoints corresponds to a Poynting flux whose axial component
gives the energy flux entering the system from the $z$-boundaries
$S_z = B_0 \mathbf{u_s} \cdot \mathbf{b}$
\citep[e.g., see][]{2008ApJ...677.1348R},
where $\mathbf{u_s}$ is the velocity at the $z$-boundary and
$\mathbf{b}$ the magnetic field at the $z$-boundary.
Because the characteristic $z$-boundary velocity timescales
are much longer than the Alfv\'en transit time $\tau_A$,
initially $S_z$ grows linearly in time akin to $\mathbf{b}$.
But once the dynamics transition to a fully turbulent regime
the system attains a statistically steady state where
the Poynting flux is on average balanced by energy dissipation,
so that also velocity and magnetic field saturate fluctuating around
their mean values.

\begin{figure}
\begin{centering}
\includegraphics[width=1.\columnwidth, bb = 150 40 635 500]{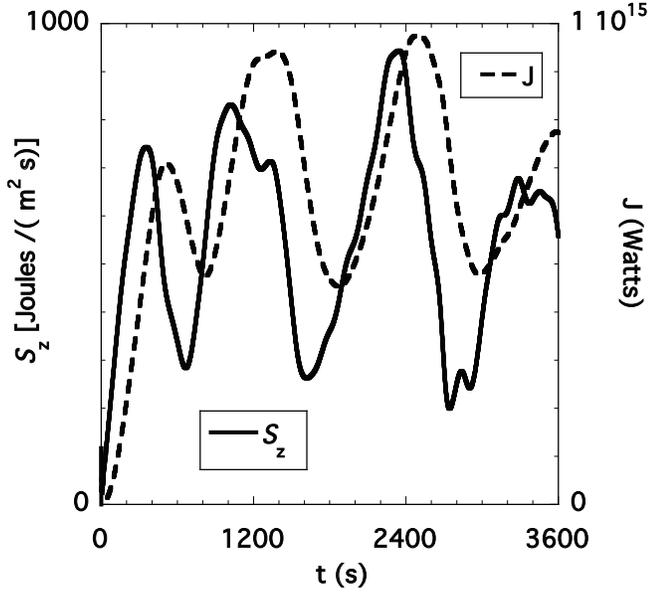}
\caption{Poynting flux ($S_z$) and Joule heating (J) vs.\ time (t) for case A.
\label{h2}}
\end{centering}
\end{figure}

Figure~\ref{h2} shows the Joule heating and Poynting flux in dimensional form
as functions of time for case~A, integrated respectively over the entire volume and over
both $z$-boundary surfaces.  Akin to our previous reduced MHD simulations
the Poynting flux exhibits large fluctuations about its average value.
This occurs because the Poynting flux contains the scalar product of the velocity at
the $z$-boundary $\mathbf{u_s}$, a given quantity, and the perpendicular
component of the magnetic field $\mathbf{b}$ that is determined by the
nonlinear turbulent dynamics of the system.  This input energy flux is therefore also a
turbulent quantity with large fluctuations in time.
Note that because the $z$-boundary velocity field
changes only slowly in time, the correlation between the velocity and the magnetic
field at the $z$-boundaries is always strong so that the Poynting flux is
always positive, i.e., energy is never removed from the loop by the boundary motions
\cite[for a study of the correlation between boundary velocity
and magnetic field see][]{2010ApJ...722...65R}.
For the latter to occur, the $z$-boundary velocity field should change over
time-scales comparable to or faster than the Alfv\'en transit time along the
loop.

In addition, the random forcing of the kind we employ is not conducive
to the formation of loop structures capable of storing a large amount of
energy.  Our forcing does not inject a net magnetic helicity -
associated with inverse cascades and therefore potentially large energy storage -
into our loop. With our forcing, the injected energy is significant enough to power a
hot corona, but clearly not a major solar flare: most of the injected Poynting flux is
efficiently converted into thermal energy as well as kinetic energy (the remaining
injected energy persists as magnetic energy of the perturbed field).

\begin{figure}
\begin{centering}
\includegraphics[width=1.\columnwidth, bb = 150 40 635 500]{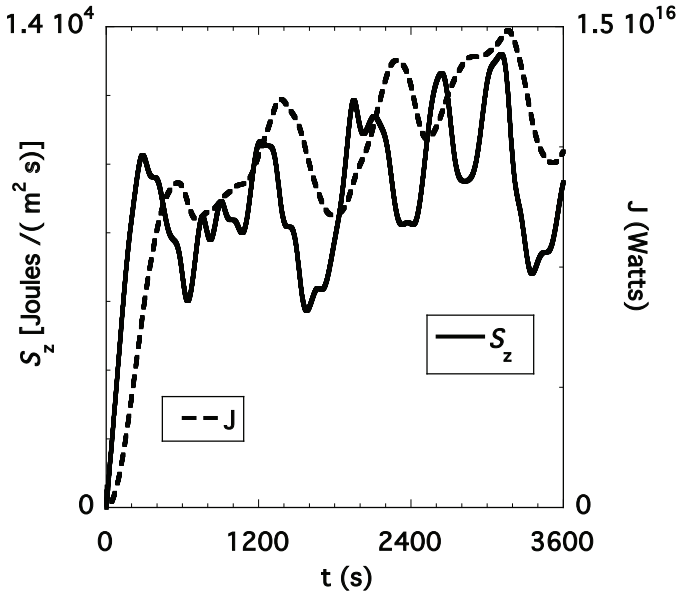}
\caption{Poynting flux ($S_z$) and Joule heating (J) vs.\ time (t) for case C.
\label{h2b}}
\end{centering}
\end{figure}

Previous reduced MHD investigations have shown that the time-averaged
Poynting flux varies approximately quadratically with the strength
of the guide magnetic field ($B_0$)
\citep[e.g.,][]{2007ApJ...657L..47R, 2008ApJ...677.1348R}.
Figure~\ref{h2b} shows the Joule heating and Poynting flux in dimensional form
as functions of time for case~C, integrated respectively over the entire volume and over
both $z$-boundary surfaces.  Recall that $B_0$ = 0.01 Tesla for Case A and
$B_0$=0.04 Tesla for Case C.
From Figure~\ref{h2} it is seen that the Poynting flux is of order
$5\times10^2$\,J~m$^{-2}$~s$^{-1}$ for Case A.
From Figure~\ref{h2b} it is seen that the Poynting flux is of order
$1\times10^4$\,J~m$^{-2}$~s$^{-1}$  for Case C.
Hence the Poynting flux increases by a factor of about twenty as the
guide magnetic field increases by a factor of four,
which within the error due to the short duration of the simulations
is consistent with a quadratic relation.

Figure~\ref{h2} also shows the Joule heating as a function of time for case A.
It can be seen that the Joule heating is somewhat correlated with the
Poynting flux -- it exhibits the same pattern of relative maxima and minima but
with a time lag (in Case A this time lag is about 200
seconds).   This lag represents the time it takes the energy to propagate in to the loop and
for the appropriate magnetic structure, i.e., electric current sheets,  to form to permit
dissipation.  Similar remarks apply to Figure~\ref{h2b} that considers Case~C.

\subsection{Three-dimensionality and intermittency}

The fluctuations seen in the Joule heating in Figure~\ref{h2}
are also evidence of temporal intermittency.
Although the numerical simulations presented here have a relatively
low spatial resolution they do present some level of intermittency.
As expected, and as shown for this problem in our previous reduced
MHD simulations, both temporal and spatial intermittency increase
at higher resolutions, i.e., with Reynolds number
\citep{2008ApJ...677.1348R, 2010ApJ...722...65R, 2013ApJ...771...76R}.
Evidence of spatial intermittency for current density and temperature
was already shown in our previous fully compressible simulations\citep{2012A&A...544L..20D}
It was found that temperature is not uniform in space,
rather it strongly increases in and around electric current sheets, forming
similarly shaped spatial structures elongated in the direction
of the strong guide field $B_0 \mathbf{\hat{e}}_z$
\citep{2012A&A...544L..20D}.
Note that both temporal and spatial intermittency should increase as the Lundquist numbers increase.

\begin{figure}
\begin{centering}
\includegraphics[width=1.\columnwidth, bb = 150 40 635 500]{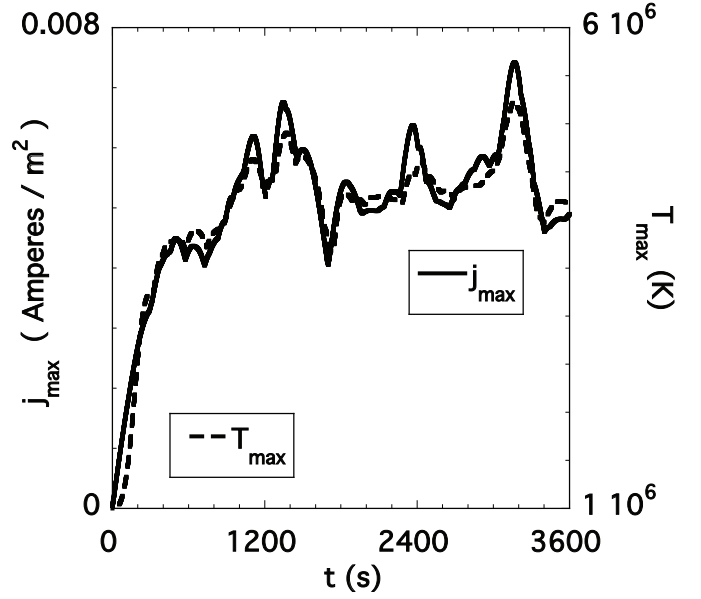}
\caption{Temperature maximum ($T_{max}$) and current maximum ($j_{max}$) {\it vs} time (t) for run C.
\label{h3}}
\end{centering}
\end{figure}

Our 3D compressible MHD simulations
allow exploration of  some of the thermodynamic implications of this turbulent
and intermittent type of heating.
The coronal loop in our simulation is in a self-consistent state, energetically determined by
the balance between boundary-forcing, nonlinear dynamics, heating and cooling.
To this must be added the non-trivial caveat that the energy flux entering the system is not
determined simply by the $z$-boundary velocity $\mathbf{u_s}$,
but also by the nonlinear, turbulent dynamics developing in the loop, 
This is a consequence of 
the Poynting flux being given by the scalar product between the $z$-boundary
velocity and the orthogonal magnetic field component generated by the
nonlinear dynamics $S_z = B_0 \mathbf{u_s} \cdot  \mathbf{b}$.
The heating is \emph{only} due to  resistive and viscous dissipation
which happens at different locations at
different times where small scales are produced, i.e., within current sheets continuously
forming and disrupting.
The behavior of the volume-averaged quantities, such as kinetic and fluctuating
magnetic energies and resistive and viscous dissipation show a temporal behavior
similar to previous RMHD results
\citep[e.g.,][]{2007ApJ...657L..47R, 2008ApJ...677.1348R, 2010ApJ...722...65R}.

Fully compressible simulations with HYPERION show the time evolution of
the maximum electric current, as seen in Figure~\ref{h3} for case D, which shows
already some fluctuations.
Figure~\ref{h3} also shows the maximum temperature $T_{max}$ as a function
of time, which correlates strongly with $j_{max}$. Though not shown here, this
correlation is seen to strengthen in simulations where the axial magnetic field strength is
increased. Indeed, increasing the axial field brings
our 3D MHD simulations closer in nature to the RMHD case.

\begin{figure}
\begin{centering}
\includegraphics[width=1.\columnwidth, bb= 140 40 610 500]{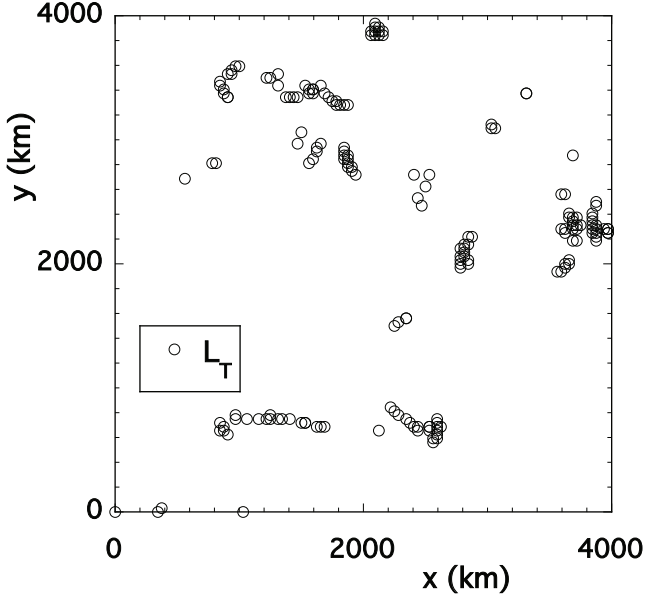}
\caption{The $x$ and $y$ maximum temperature locations for run D.
The term $L_T$ denotes the location of the temperature maximum.
\label{h4}}
\end{centering}
\end{figure}

\cite{2012A&A...544L..20D} showed that the temperature is
spatially structured, i.e., it is spatially intermittent.
Figure~\ref{h4} shows the $x$ and $y$ positions of $T_{max}$ in space for case~D
at selected times. It can be seen that $T_{max}$ wanders about,
observationally resulting in a changing radiation emission pattern
that can easily give the mistaken impression of an oscillating loop.

\begin{figure}
\begin{centering}
\includegraphics[width=1.\columnwidth, bb= 100 85 620 455]{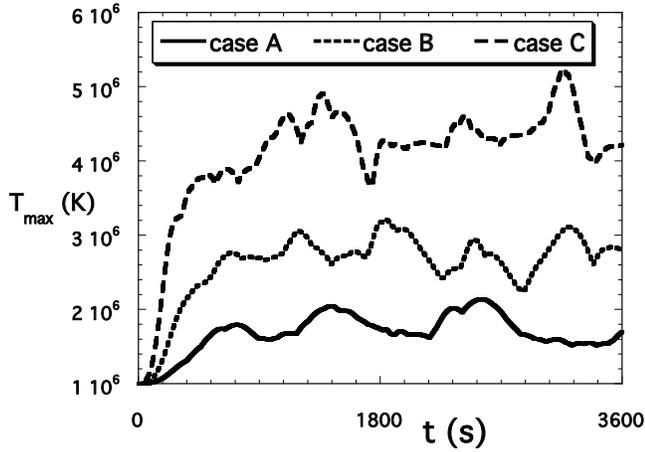}
\caption{Comparison of maximum temperature versus time for cases A, B, and C.
\label{h5}}
\end{centering}
\end{figure}

As seen in Figure~\ref{h1}, there is considerable variation in loop lengths
and magnetic field strengths in the solar corona.
Here we briefly consider the influence of the axial magnetic field strength
on our results.
Figure~\ref{h5} shows how the maximum temperature depends on the axial magnetic field
strength (cases A, B, and C).
It can be seen that the maximum temperature increases with the magnetic
field strength, with a slightly weaker than linear dependence on field strength.

\begin{figure}
\begin{centering}
\includegraphics[width=1.\columnwidth, bb = 100 85 620 455]{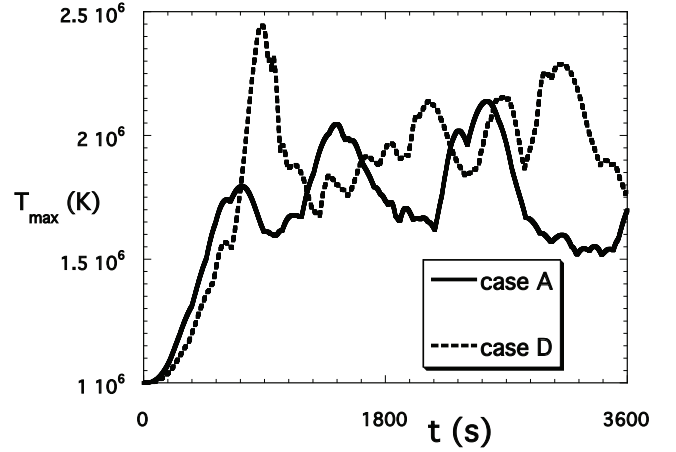}
\caption{Comparison of maximum temperature {\it vs} time for cases A and D.
\label{h6}}
\end{centering}
\end{figure}

How do the results change with horizontal ($x$ and $y$) resolution and Lundquist numbers?
Case D has twice the horizontal resolution as Case A.
In addition, the Lundquist numbers are doubled and the Prandtl
numbers are halved for Case D.
Figure~\ref{h6} shows how the maximum temperature depends on numerical resolution and
the Lundquist numbers (cases A and D).
Note that the RMS temperatures are not too different for the two cases, but the
temperature oscillations in the higher Lundquist number case are somewhat stronger.
Of course as in all turbulent systems, the full understanding of the high
Reynolds number regime is non trivial, and it will be investigated
in future work.
Nevertheless our previous reduced MHD simulations indicate
that dissipation rates and Poynting flux saturate at resolutions
of about $256^2 \times 128$, and as shown here maximum temperature
variation is weak at $128^2 \times 144$.

\subsection{Effects of vertical ($z$) numerical resolution}
The energization and response of the system depend on gradients at the $z$ boundaries \citep{2011ApJS..194...26B, 2013ApJ...773...94}.
The most significant are the gradients of the magnetic vector potential, the temperature, and the number density.
The Poynting flux depends on the magnitude of the $x$ and $y$ magnetic fields.
In HYPERION these fields are computed as the curl of the magnetic vector potential.
For the $x$ and $y$ magnetic fields there is a component due to the $z$ gradients of the $y$ and $x$ magnetic vector potential,
hence the energization of the system depends on the accurate computation of these gradients.
In the same way the response of the system to heating depends on the evolution of thermodynamic gradients near the $z$ boundaries.
At first glance it might appear that we have under-resolved these gradients.
The scale height of the initial temperature [$T_i / (dT_i / dz)$] can be estimated in nondimensional terms from equation~\ref{eq:tempinit}.
At the $z$ boundaries the temperature scale length is found to be 0.00789 (31.56 km in dimensional terms).
For our system with $L_z = 12.5$ (or 50000 km), this is resolved using 1585 uniformly spaced mesh points.
Note that the initial number density scale height will be approximately the same.
We will determine {\it a~posteriori} how the $z$ resolution affects the energization and plasma response.
Case A will be used as the baseline.  In these simulations all of the physical parameters are the same as in case A; only the $z$ resolution is changed.
Case A has 144 points in $z$, case E has 288 points, case F has 576 points and case G has 1620 points (sufficient to resolve the temperature and
number density scales at the boundaries).
All of these cases have a dimensional magnetic field of 0.01 Tesla, so the stiffening effect of the DC magnetic field is
weak relative to the other runs.
Case G was only simulated for approximately 1800 seconds to allow for the computation of synthetic emissions and an emission measure, to be
shown in a subsequent part of this paper.

\begin{figure}
\begin{centering}
\includegraphics[width=1.\columnwidth, bb = 100 85 620 455]{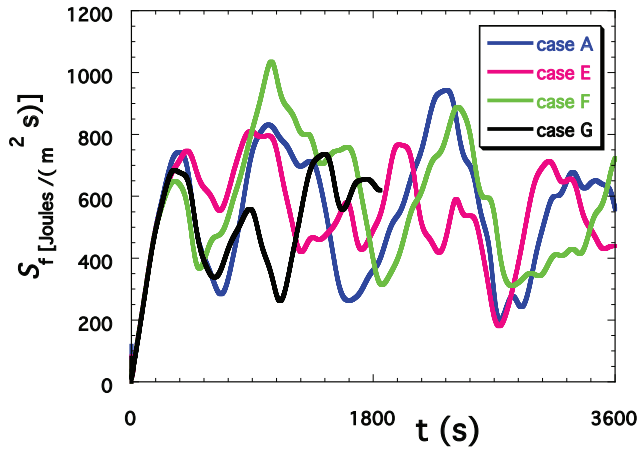}
\caption{Comparison of Poynting flux {\it vs} time for cases A, E, and F.
\label{pflux_aefg}}
\end{centering}
\end{figure}

A comparison of the Poynting flux for the simulations with different $z$ resolutions is shown
in figure \ref{pflux_aefg}.   It can be seen that all of the cases oscillate about approximately the same average value in time.
In HYPERION the magnetic vector potential is advanced in time.
Recall that the Poynting flux depends on the perpendicular component of the magnetic field.
Hence the perpendicular magnetic field is a derived quantity -- in particular
it will depend on $z$ derivatives.  The value of these derivatives will vary somewhat with the $z$ resolution.  The nonlinearity of
the system is reflected in the temporal variability shown in figure ~\ref{pflux_aefg}.

\begin{figure}
\begin{centering}
\includegraphics[width=1.\columnwidth, bb = 100 85 620 455]{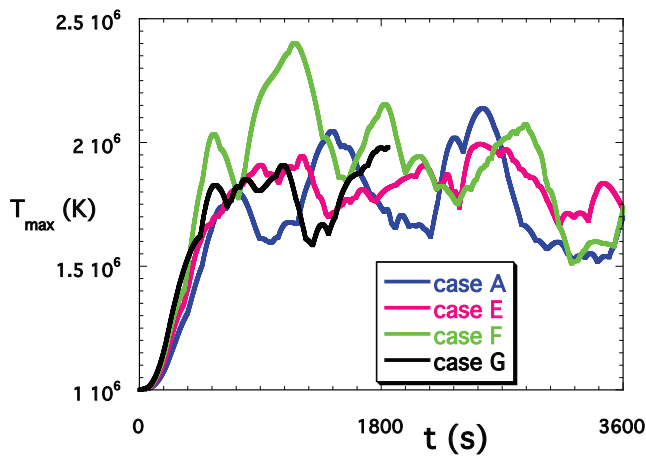}
\caption{Comparison of maximum temperature {\it vs} time for cases A, E, and F.
\label{tmax_aefg}}
\end{centering}
\end{figure}

A comparison of the maximum temperatures for the simulations with different $z$ resolutions is shown
in figure \ref{tmax_aefg}.  The behavior here is similar to that exhibited by the Poynting flux.  The maximum temperature
oscillates about approximately the same value for all of the cases, with some variability seen in the details of the fluctuations.
We conclude that, for the range of $z$ resolutions we have considered here, there is not a significant change in the numerical results.

\subsection {Emission measure distribution}

\begin{figure*}
\center
  \includegraphics[height=.65\textwidth]{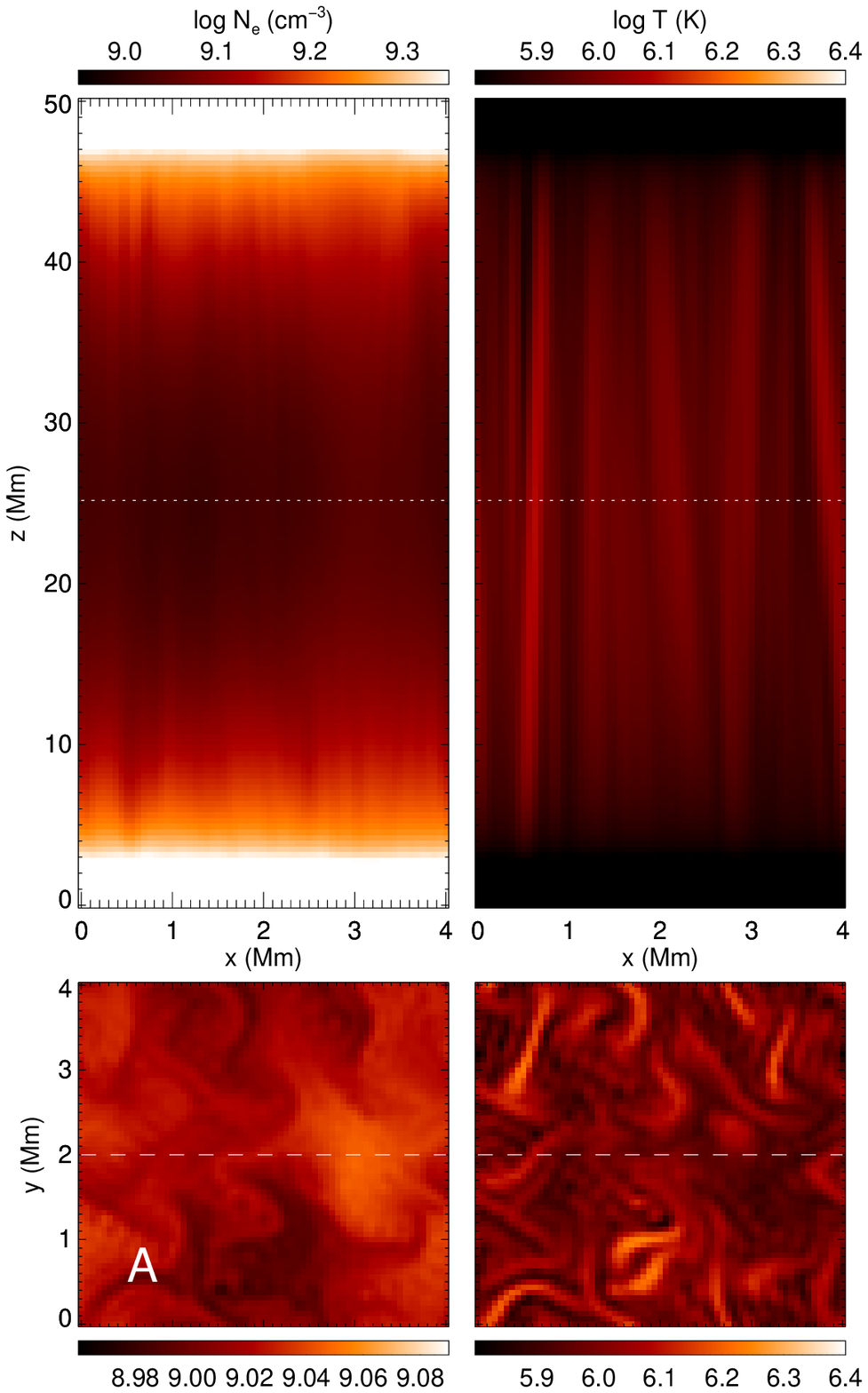}
  \includegraphics[height=.65\textwidth]{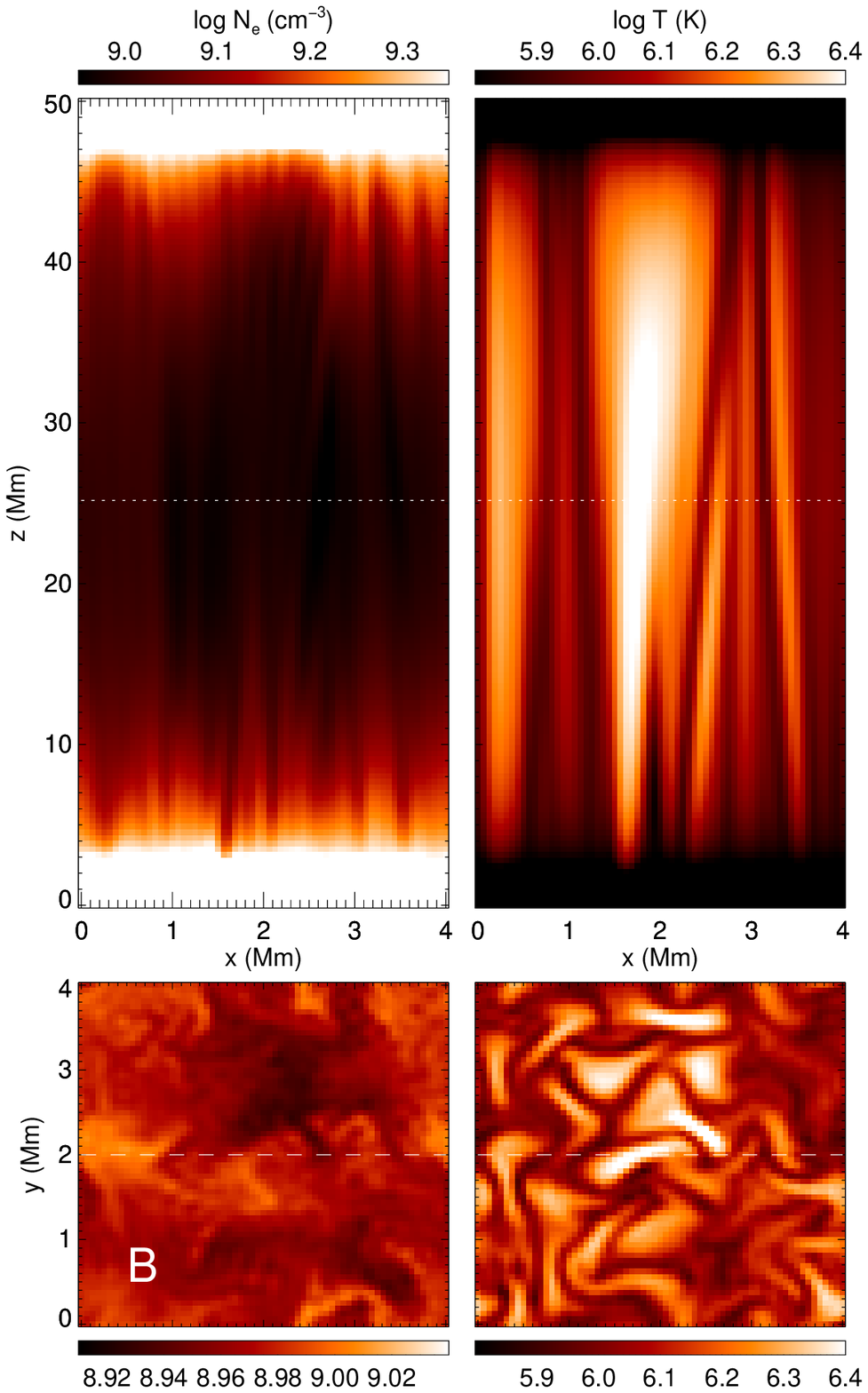}
  \includegraphics[height=.65\textwidth]{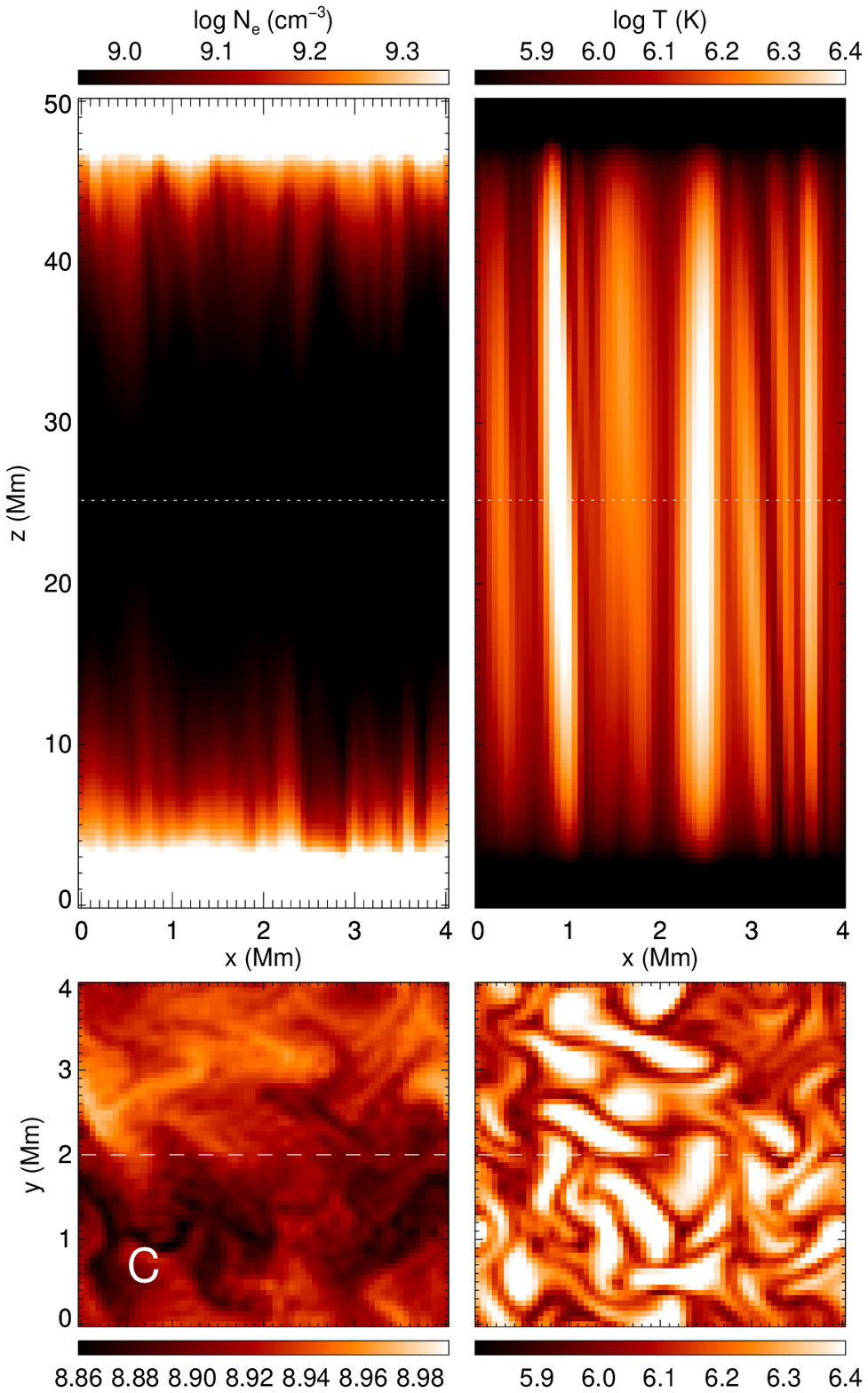}
  \includegraphics[height=.65\textwidth]{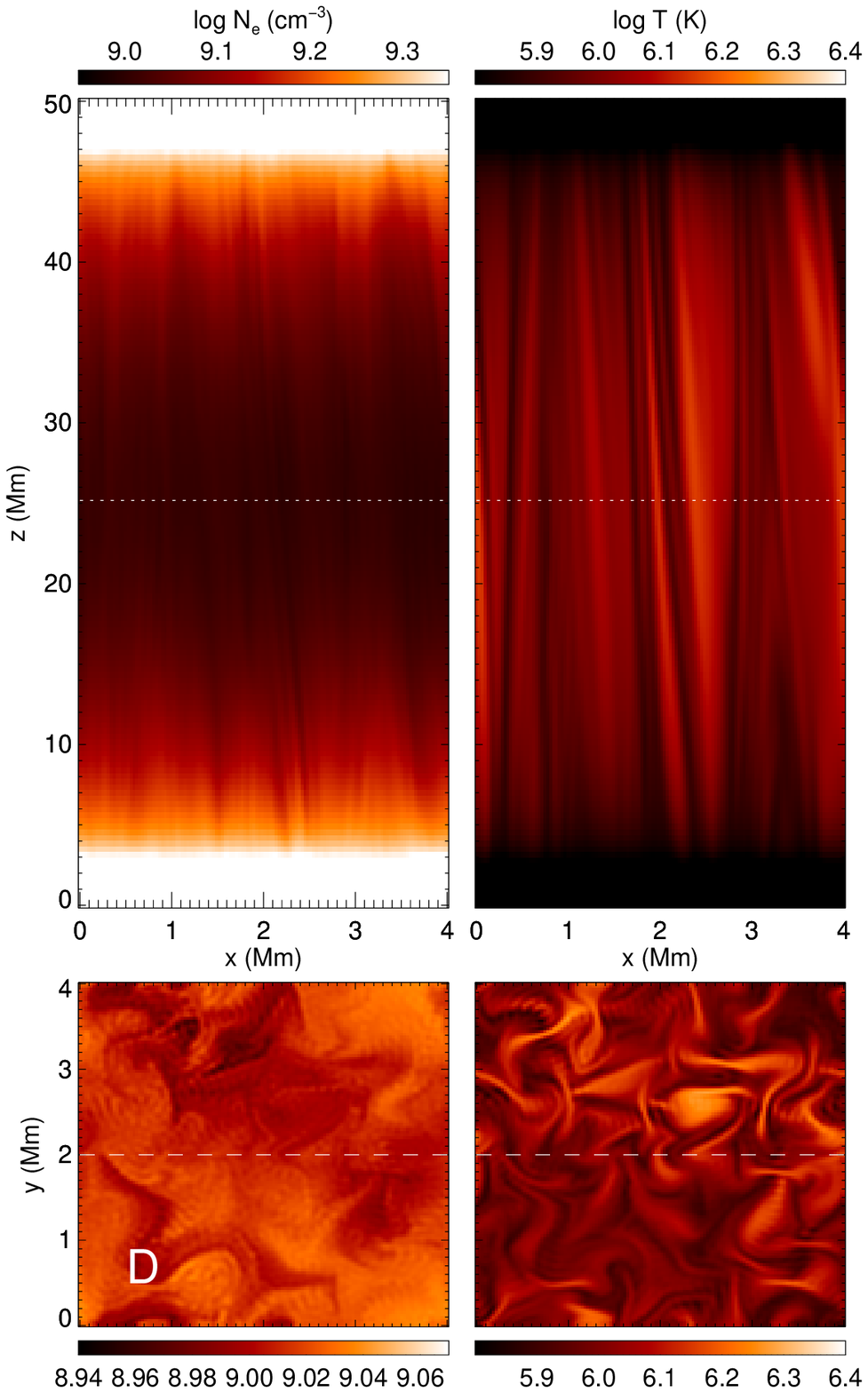}
 \caption{ Temperature and density distributions along and across the
   loops for all cases A through D.
  \label{h7}}
\end{figure*}

\begin{figure*}
  \centerline{\includegraphics[bb=0 26 595
      350,clip,width=0.9\textwidth]{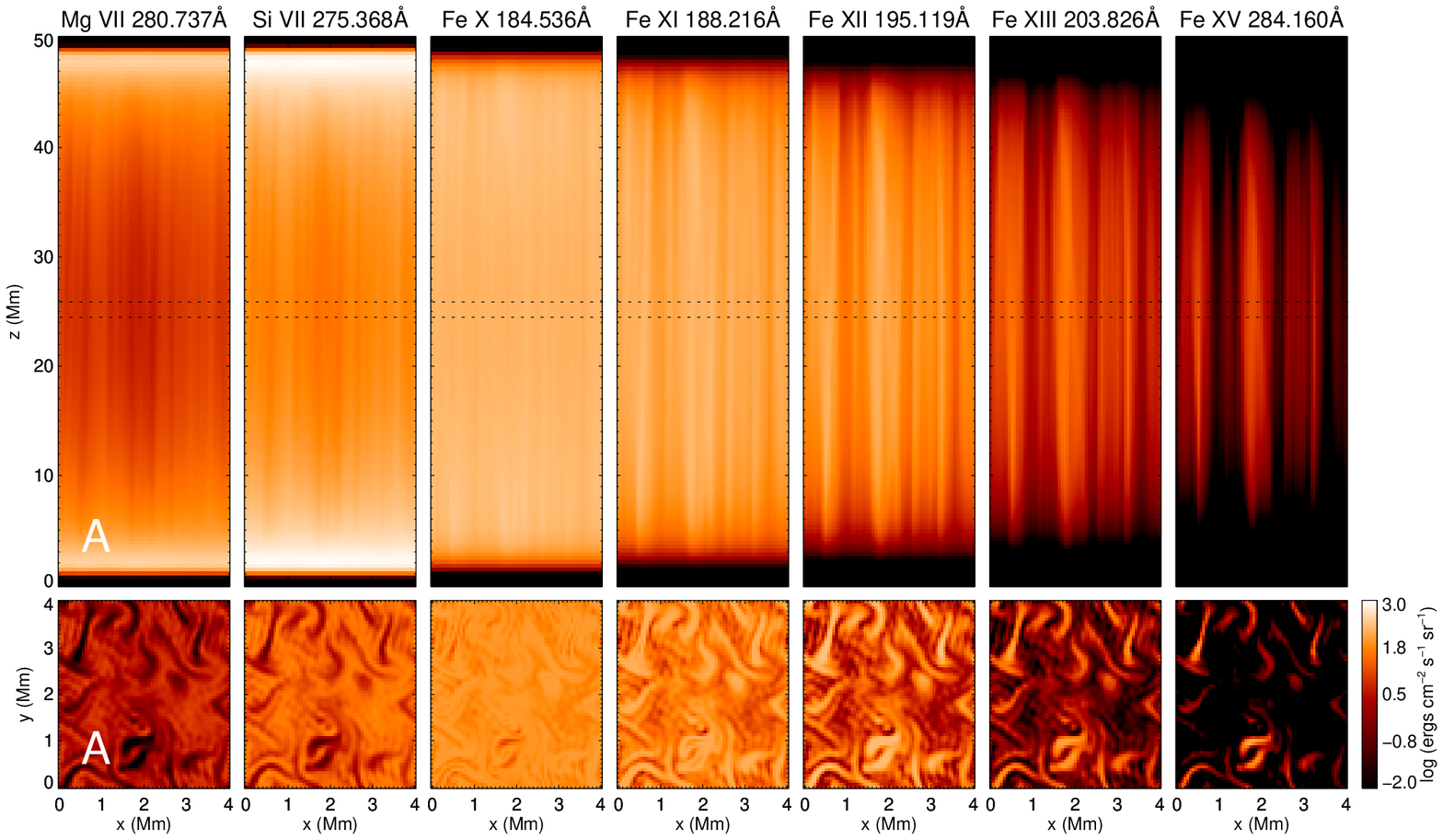}}
  \centerline{\includegraphics[bb=0 26 595
      116,clip,width=0.9\textwidth]{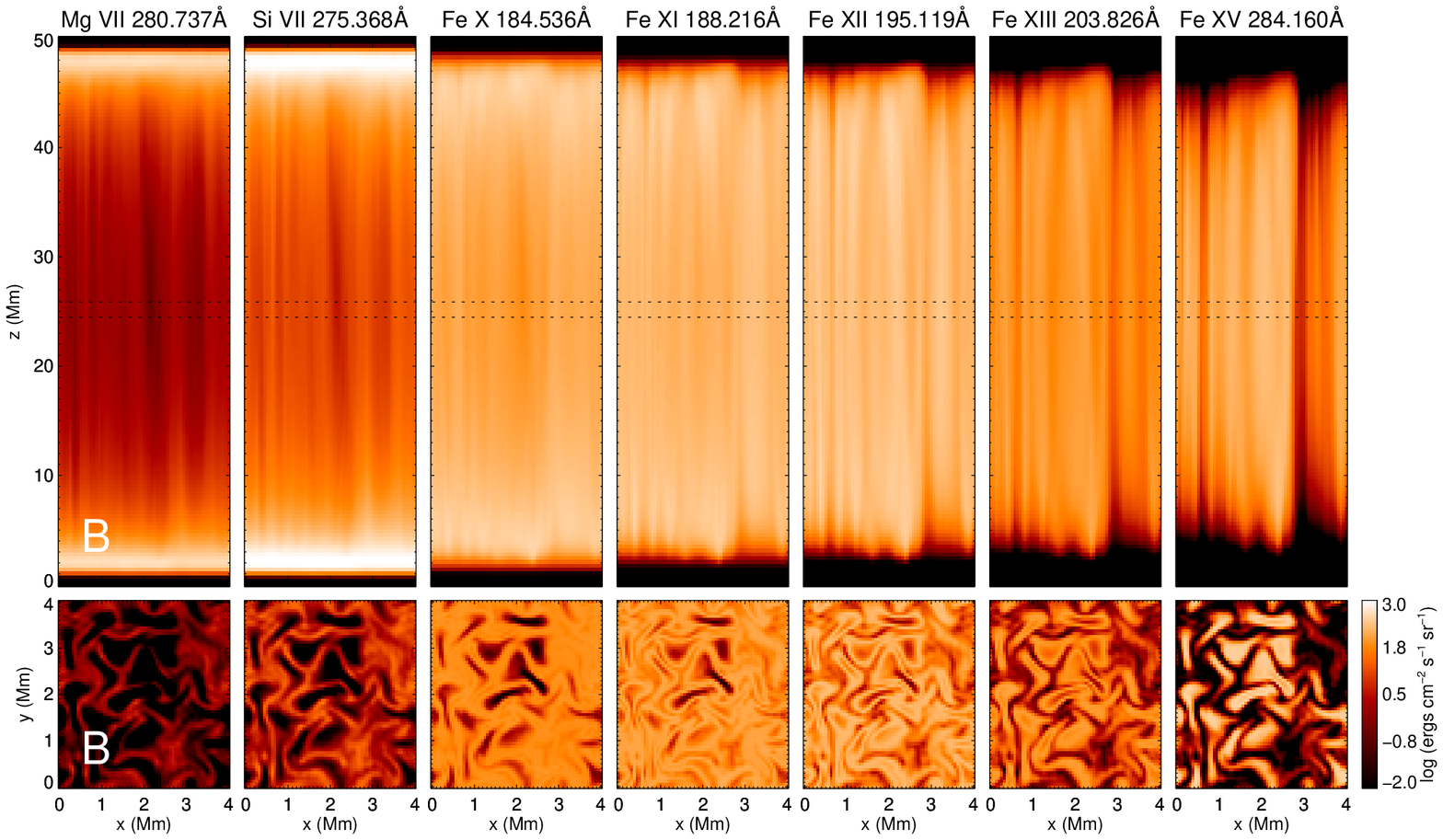}}
  \centerline{\includegraphics[bb=0 26 595
      116,clip,width=0.9\textwidth]{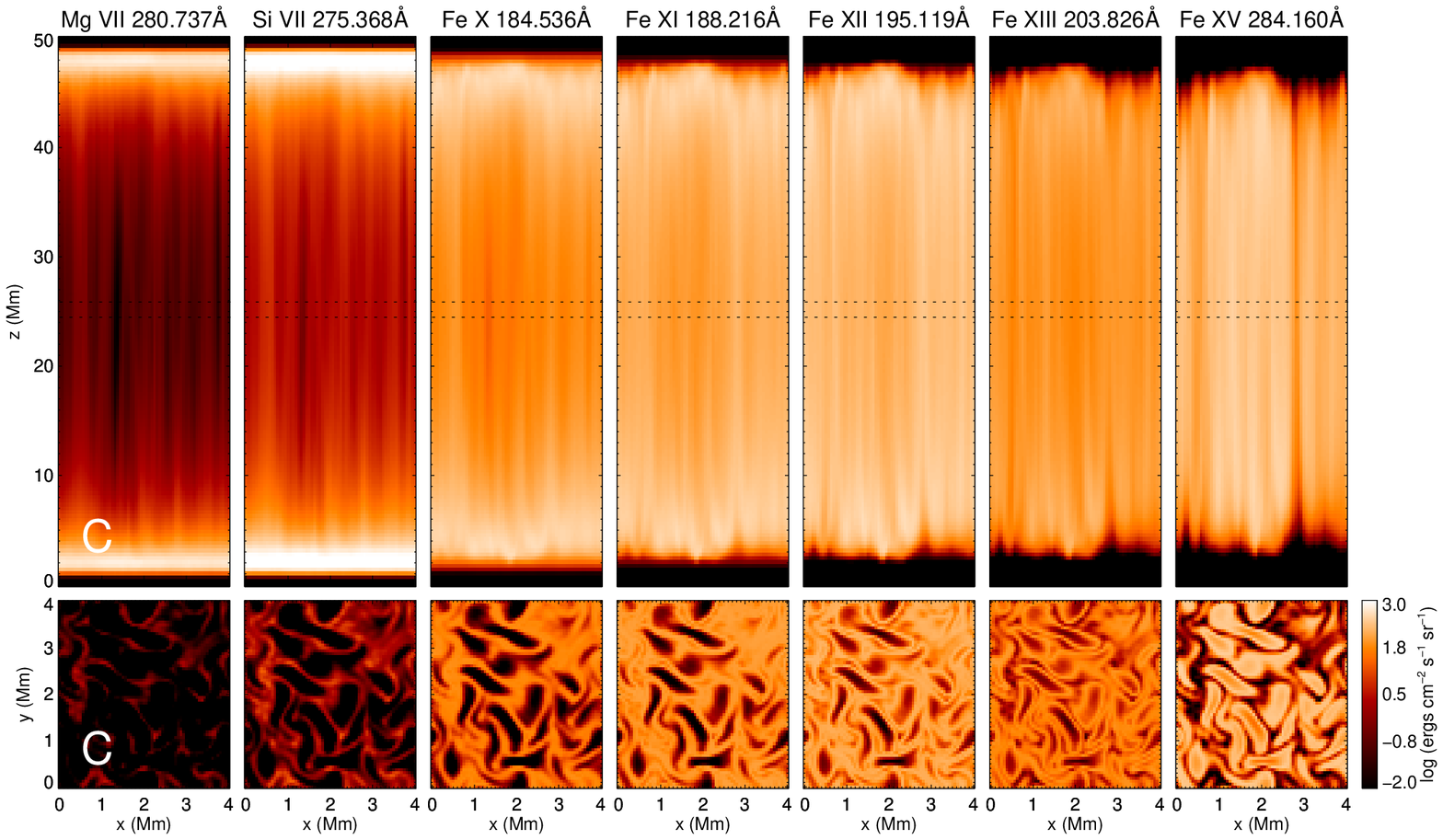}}
  \centerline{\includegraphics[bb=0 15 595
      116,clip,width=0.9\textwidth]{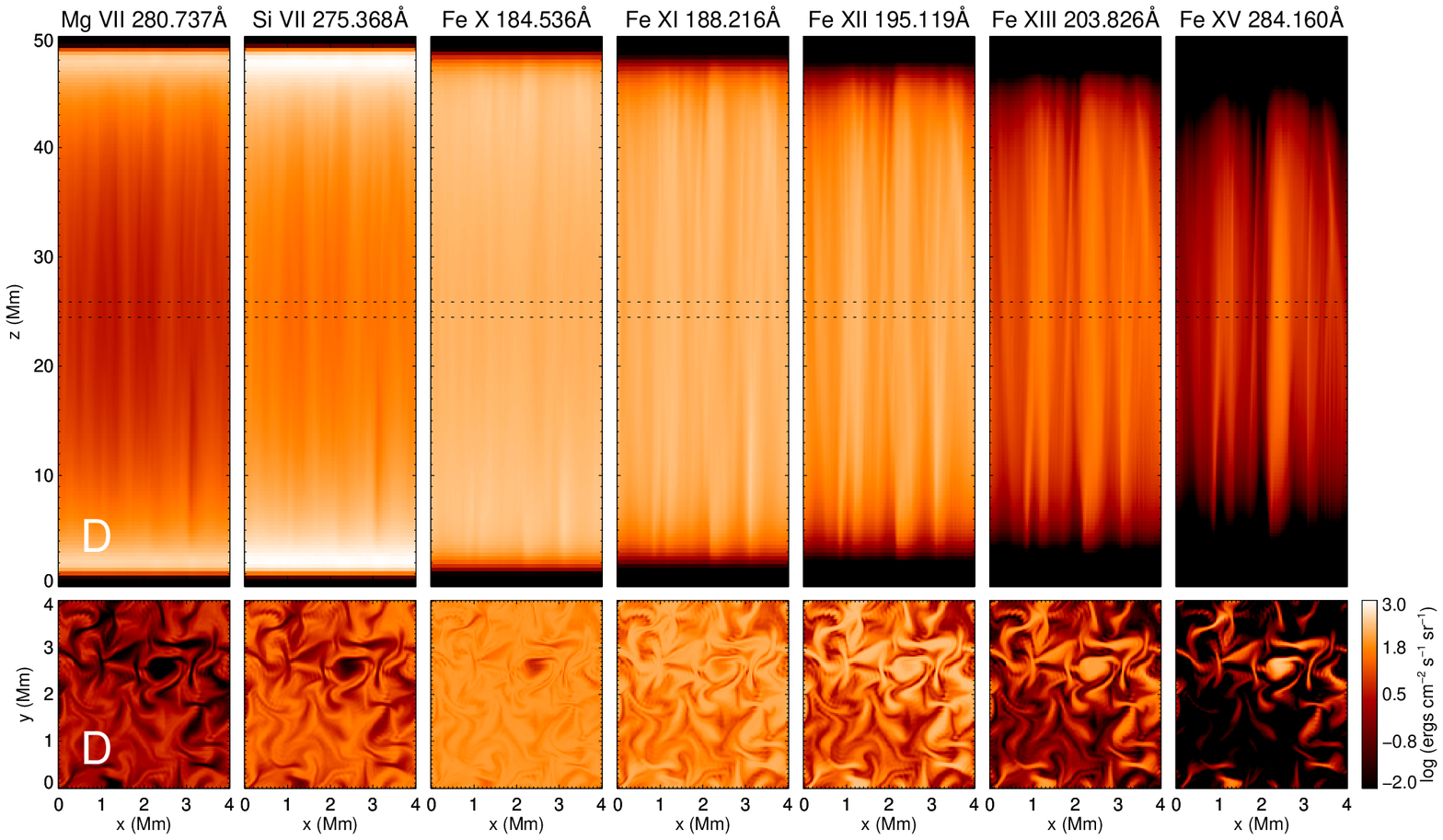}}
  \centerline{\includegraphics[bb=0 115 595
      340,clip,width=0.9\textwidth]{figures/sum_int_009a.eps}}
\caption{Synthesized intensities along and across the simulation box for a subset of the spectral lines used for the emission measure analysis. Middle panels show intensities integrated along five voxels in the $z$ direction. Top and bottom are integrations along the full range in $y$. The same scaling and units apply to all of the panels.
\label{h9}}
\end{figure*}

Since an emission line is formed over a relatively narrow temperature
range, spectrally resolved observations can be used to infer the
temperature structure of the solar atmosphere. This is often achieved by
computing the differential emission measure distribution (DEM), which is
a solution to the equation
\begin{equation}
 I_{i} = \frac{1}{4\pi}\int \epsilon_i(T)\xi(T)\,dT.
\end{equation}
Here $I_{i}$ and $\epsilon_i(T)$ are the intensity and emissivity of the
emission line. The emissivity includes all of the information specific
to the atomic transition.  The quantity $\xi(T)$ is the DEM, which
describes the conditions in the solar atmosphere, and is written
\begin{equation}
 \xi(T) = n_e^2\frac{ds}{dT},
\end{equation}
where $n_e$ is the electron density and $s$ is a coordinate along the
line of sight.  Further details and an application to solar observations
can be found, for instance, in \cite{2008ApJ...686L.131W}.

The density and temperature for each voxel, that is each element of the
simulation volume, were used
to calculate the intensity of EUV spectral
lines. We chose a set of 25 EUV lines ranging from $3\times10^5$ K to
$7\times10^6$ K in temperature of formation.  With the exception of
\ion{Fe}{18} 974.86 \AA, the lines selected are all in the observed
wavelength range of the EIS instrument on board Hinode
\citep{2007SoPh..243...19C} and cover a variety of ionization stages of
Mg, Si, Fe, S, Ar and Ca. Data from the EIS instrument have been
routinely used to calculate emission measure distributions in different
coronal conditions. The \ion{Fe}{18} 974.86 \AA\ was added to improve
the constraints on the high temperature end and mimics the use of AIA
94\,\AA, which images \ion{Fe}{18} \citep{2012ApJ...759..141W}. The
emissivities for each line was calculated using the CHIANTI atomic
database \citep{1997A&AS..125..149D, 2013ApJ...763...86L} assuming
coronal abundances \citep{1992ApJS...81..387F} and the CHIANTI
ionization equilibrium tables.

Figure~\ref{h7} shows the density and temperature distributions along
and across sections of the simulation domain, for time from 1770 s to 1830 s
for the simulated cases A through D. Figure~\ref{h9} shows at the top and bottom panels
the synthesized intensities of a set of seven spectral lines integrated
along the perpendicular direction to
the loops' axis, similar to observing the loops side-on. The panels in the
center are the intensities of the loops' mid-section integrated along five
voxels in the $z$ direction. The integration times are 60 s in all four
cases.

The emitting volume selected for the EM exercise corresponds to the apex of the loops. To compute line intensities we integrate the emissivities over a region 1750 km wide centered at the mid-plane of the computational domain. The volume of integration corresponds therefore to a 
$4000\times1750~\rm
km^2$ area on a hypothetical plane of the image, 4000 km deep, namely a
cross-section of about 5\arcsec, typical of loop observations in the
corona. The integrated intensities in each spectral line, with an assumed
uncertainty of 20\%, serve as input
to a Differential Emission Measure calculation algorithm. We used the
Monte Carlo Markov Chain (MCMC) code \citep{1998ApJ...503..450K},
applied in the manner described in \cite{2012ApJ...759..141W}.  The MCMC
algorithm calculates multiple (250) solutions with perturbed values for
the intensities, providing an estimate of the error in the EM
distribution calculation.

\begin{figure}
  \centerline{\includegraphics[width=1.\columnwidth]{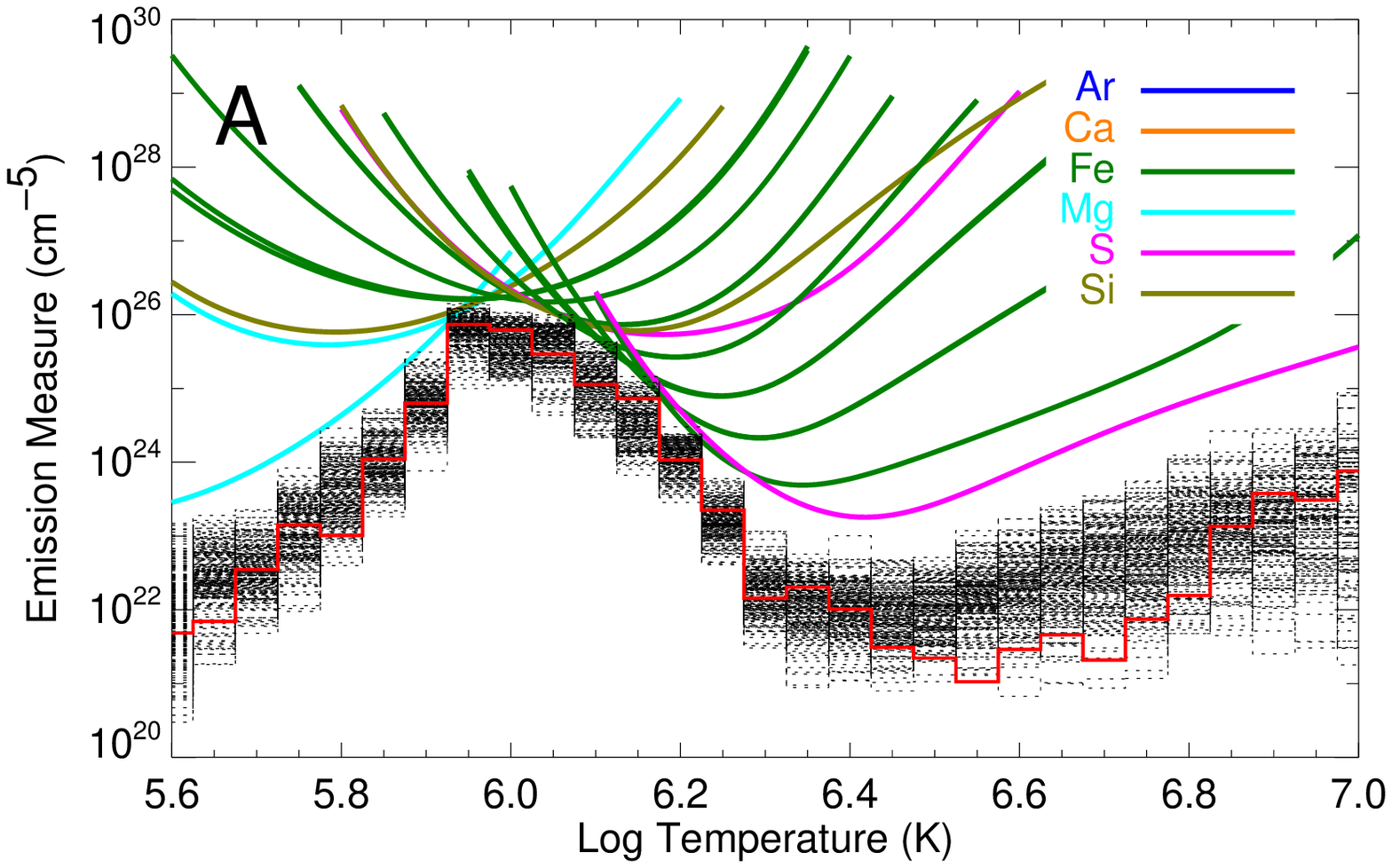}}
  \centerline{\includegraphics[width=1.\columnwidth]{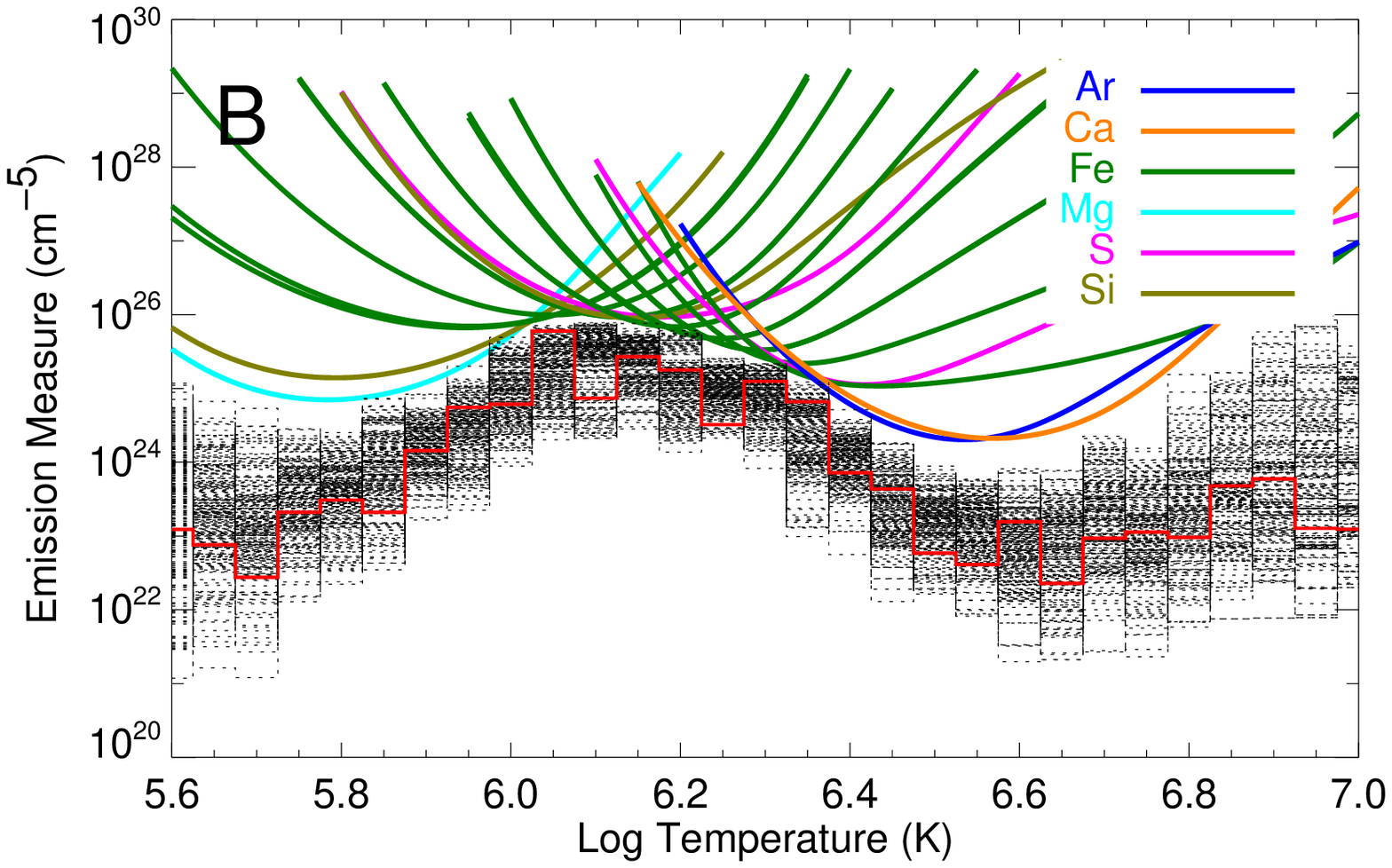}}
  \centerline{\includegraphics[width=1.\columnwidth]{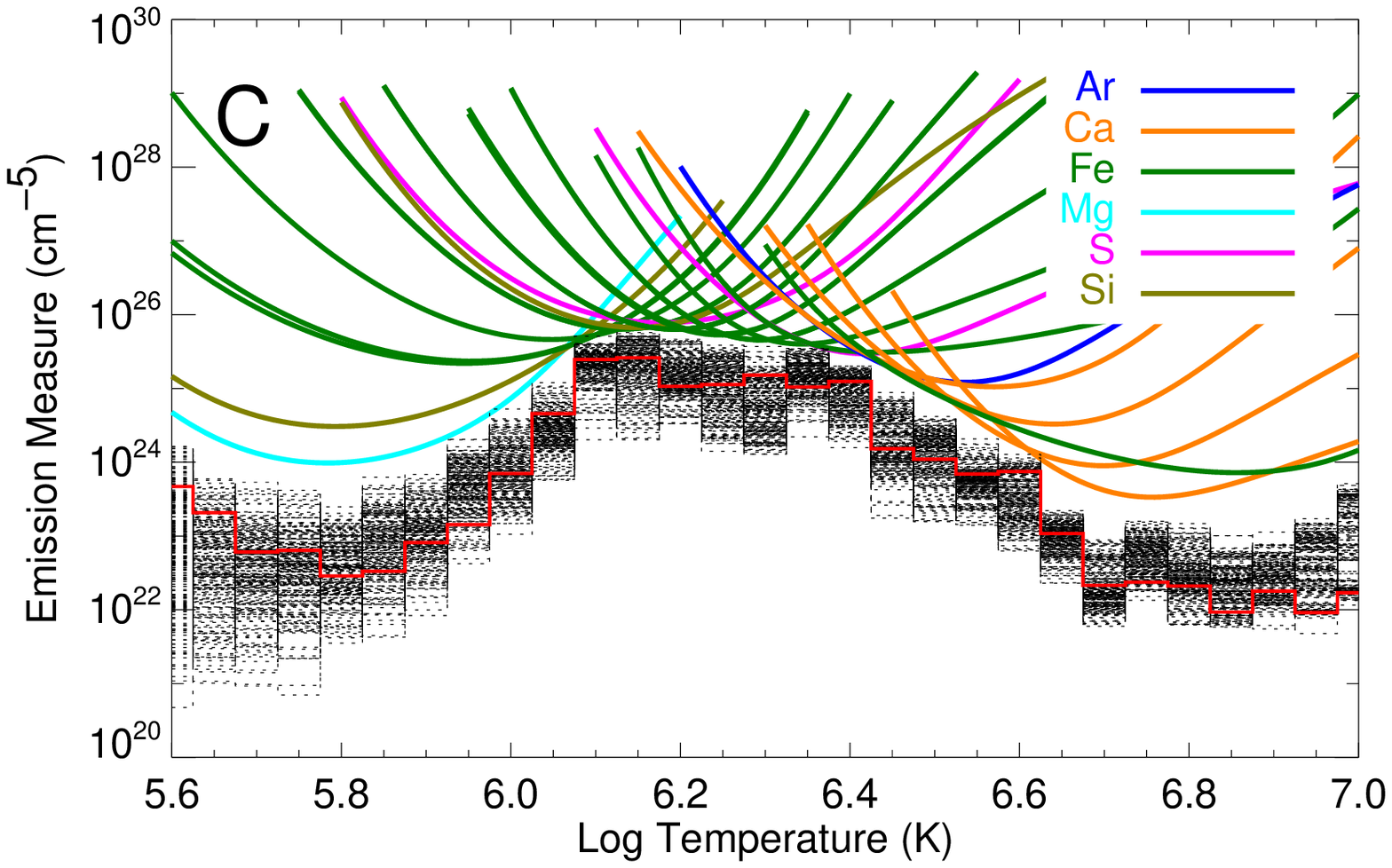}}
  \centerline{\includegraphics[width=1.\columnwidth]{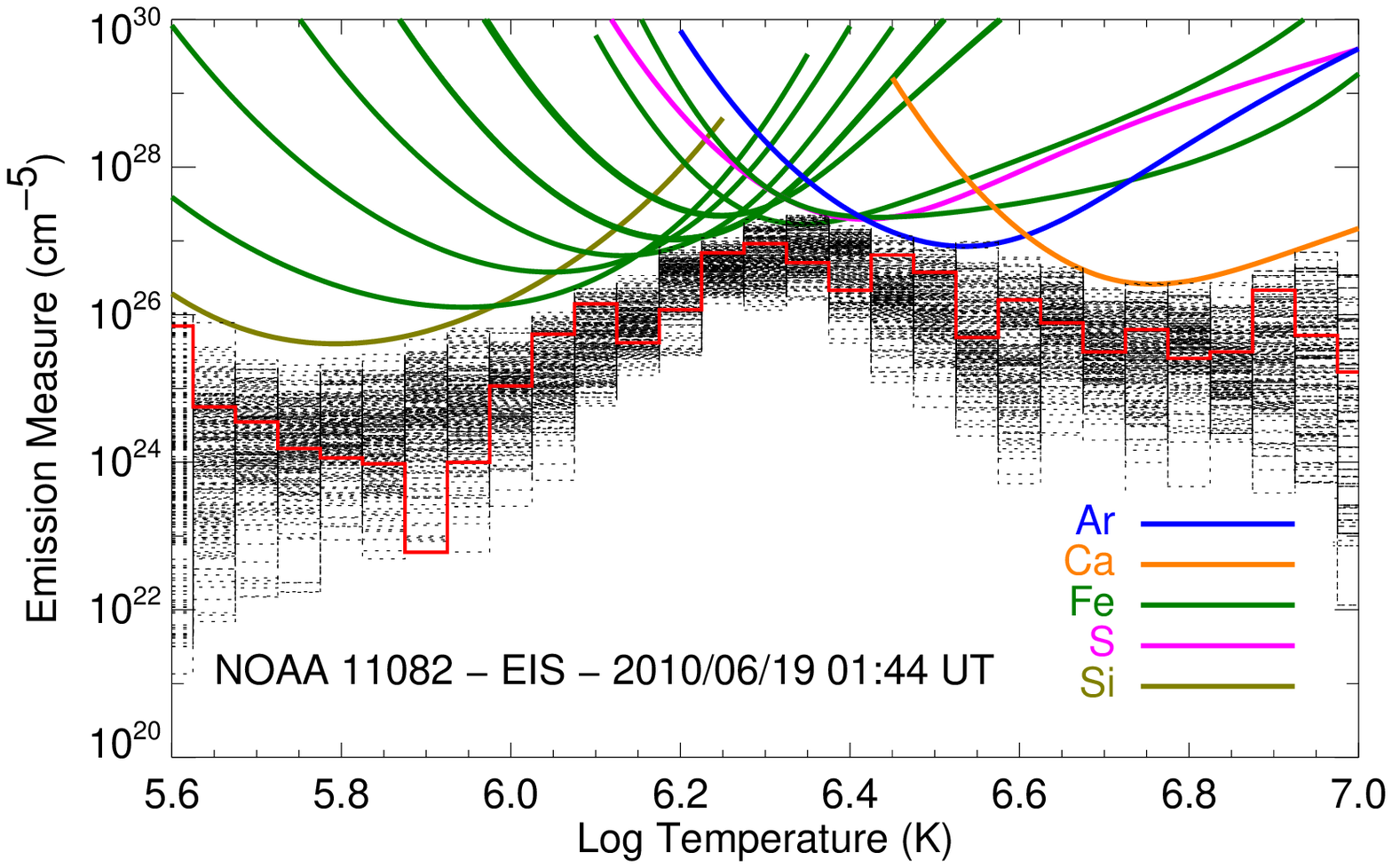}}
\caption{\label{h11}Emission measure distributions in the mid-section
  (loop apex) for cases A, B, and C. The bottom panel shows the DEM
  computed using observed intensities from the region shown in Figure~\ref{h13}.}
\end{figure}

Figure~\ref{h11} shows the EM solutions for the A, B and C runs, where the red
line corresponds to the best-fit solution.  The plot also shows, for
context, color-coded lines representing the emission measure loci for
the different atomic species of the EIS spectral lines. They illustrate
the range of temperature dependent emission measures compatible with the
intensity of a particular spectral line. A set of intensities at
different temperatures constrain the EM distribution compatible with the
complete dataset.

\begin{figure}
\centerline{
  \includegraphics[width=1.\columnwidth]{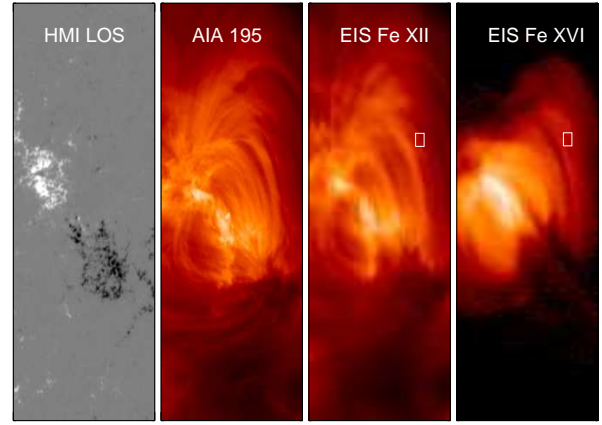}
  }
\caption{\label{h13}NOAA 11082 as seen by HMI, AIA and EIS on June 19, 2010. The box marks the region
of integration for emission measure analysis at the bottom of Fig.~\ref{h11}.}
\end{figure}

The computed emission measure distributions are qualitatively similar to
the emission measure distributions computed from observed intensities,
with a characteristic Gaussian-like distribution in the 1--4 MK
temperature range. The weighted mean temperatures for the the three
cases are respectively: 6.00, 6.13 and 6.22 MK. These are characteristic
temperatures for the spectral windows (e.g., \ion{Fe}{12}) and filter
bandpasses (171 \AA, 195 \AA) where we observe a significant fraction of
the loop emission in the corona and are consistent with the peak
emission measure temperatures of some of these loops
\citep{2008ApJ...686L.131W}.  The bottom panel of Figure~\ref{h11} shows as
a comparison of the EM distribution for a sample area in active region NOAA
11082 (Figure~\ref{h13}), illustrating the similarities of the simulation
emission measures with regions of the corona.
The EM distributions are perhaps the easiest way to compare the simulations
with the observations. The simulated intensities for individual spectral lines
can differ from typical observed values by factors of 2--10. Such line-by-line
comparisons are beyond the scope of this work, but will be considered in the future.
Temperatures can be higher at the core of
active regions, with emission measure distributions peaking at $\sim$4
MK and exhibiting asymmetric profiles with a steeper drop in the high
temperature end \citep{2012ApJ...759..141W}.

\begin{figure}
\centerline{
  \includegraphics[width=1.\columnwidth]{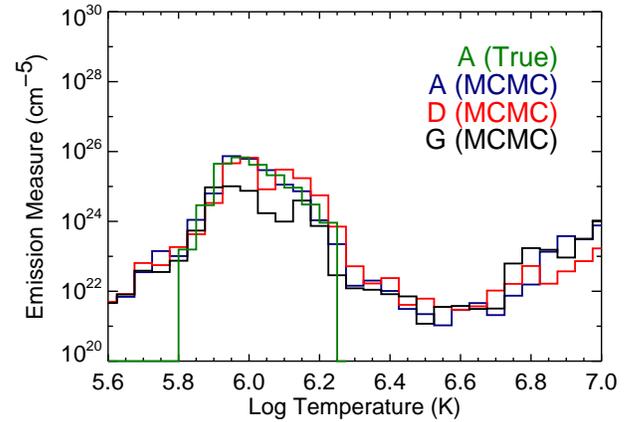}
  }
\caption{\label{h12}A comparison of the true emission measure
  and the best fit solution from MCMC for case A (Figure~\ref{h11}).
  Low and high resolution runs (cases A, D and G) exhibit
  nearly identical emission measures.}
\end{figure}

Fig.~\ref{h12} shows that the emission measure analysis is able to
restore the true line-of-sight emission measure, that is the true density
distributions as a function of temperature in the volume of integration ($\sum n_{e_i}^2 V_i/Area_{int}$).
This is applicable for the four cases A through D which are shown. The figure also
demonstrates that we do not find significant differences in the EM distribution
between the low and high resolution runs.

\section{Discussion and conclusions}

In this paper we have examined the dynamics of a coronal loop threaded by a
strong axial magnetic field, where the field line footpoints
are advected by random motions.
Consistent with previous two-dimensional and 3D reduced MHD simulations
\citep{1996ApJ...457L.113E, 1997ApJ...484L..83D, 2003PhPl...10.3584D,
2007ApJ...657L..47R, 2008ApJ...677.1348R, 2011PhRvE..83f5401R},
and our previous fully compressible work \citep{2012A&A...544L..20D},
the loop dynamics are found to be nonlinear, with a turbulent
MHD cascade transporting the energy injected by boundary motions
at large perpendicular scales to smaller scales. This leads to
the formation of approximately field-aligned current sheets
where energy is dissipated and around which temperature strongly increases,
with small scales mainly in planes perpendicular to the axial magnetic field,
along which both current and temperature structures are elongated.
These small scales are not uniformly distributed in the loop, but rather the dynamics become
increasingly more intermittent at higher Lundquist numbers both in space and in time.
Localized electric current sheets continuously form and disrupt,  leading to
localized heating of the plasma on short time scales.

Our results show that the loop is the site of a continuous occurrence of reconnection
events that present observations are unable to resolve both spatially and temporally.
In the presence of a strong guide field magnetic reconnection
occurs at the X-points of the orthogonal magnetic field
component, leading to a continuous change
of connectivity of the field lines that cross the reconnection sites (``interchange'')
where the heating occurs \citep{2007ApJ...662L.119S}.
These many sub-resolution ``heating'' events add up to produce the
observed emission,  giving the impression at larger (observational) scales
of a continuous diffuse heating.
What is called ``coronal heating'' is actually the superposition of all events due to localized
energy deposition along the subsequent different field lines that cross the reconnection sites,
at the many current sheets elongated along the axial direction present in the
loop volume at any give time
\citep[for a visualization of such current sheets, see, e.g.,][]{2008ApJ...677.1348R}.
Clearly the heating deposited along ``strands'' (small elemental flux tubes)
is much smaller than the total dissipated power in the heating peaks shown
in Figures~2--5, which is of the order of $10^{15}$ to $10^{16}$ Watts with a duration
of about 1000~s. This suggests that the energy released along each
strand is reasonably expected to be much smaller than $10^{16}$~J,
which is about $\sim 10^{-9}$ times the typical energy released in a flare.
We expect the energy deposited along strands in typical heating events
to exhibit a distribution with a peak at energy smaller than $10^{16}$~J,
and plan to investigate more in depth the energy release mechanism
and statistics in future work.

Recall that in our calculations we have used values of resistivity and viscosity that are much
larger than the real ones.  In the real Sun even smaller spatial and temporal scales are
attained leading to even smaller energies being involved in each event.  Evidence for this
has been seen in RMHD calculations \citep{2008ApJ...677.1348R, 2013ApJ...773L...2R}.
Considering the values of resistivity and viscosity we adopt are much higher
than the solar values,  we expect in each event an energy release much smaller than that for a nanoflare.
We will study this point in detail with higher resolution simulations in future.

We have employed an emission measure analysis to investigate whether the simulated
intensities of the computational loops are representative of plasma in the corona and
find great similarities both in peak temperature and distribution.
We find that the simulated  intensities and corresponding emission measures are in excellent
agreement and they are accurate representations of the true emission measures.
\cite{2012ApJ...758...54T},
looking into  3D simulations of active regions, found that this method can be inaccurate
when structures with significantly different density overlap along the same line-of-sight.
The temperatures, which increase as the value of the
axial magnetic field increases, are characteristic of warm loop
structures visible in EUV channels.
The loop is found to be a multi-temperature structure with isolated regions at temperatures
of several million degrees and most of the loop at much lower temperatures.
For each case presented in the previous sections the emission measure retains the same
form for the entire hour of the computation in spite of the strong spatial and temporal intermittency.

In this paper we have adopted a Cartesian model of a coronal loop.
The random motions at the boundaries shuffle the magnetic footpoints such
that there is no ordered twisting of the field-lines.
This random twisting does not facilitate the formation of magnetic structures
that can store large amount of energy.
Rather the system reaches a statistically steady state where integrated
physical quantities (magnetic energy, Poynting flux, dissipation rates and
radiative losses) fluctuate around their time-average values, so that
the injected energy per unit time is entirely dissipated on the average
(i.e., considering a long enough time interval).
In a Cartesian model the only way to store a large amount of energy, that can
subsequently give rise to larger magnetic energy release events, is to apply
a spatially isolated and symmetric $z$-boundary velocity.
For instance, a vortex with intensity
stronger than surrounding $z$-boundary motions can twist the coronal magnetic
field lines quasi-statically, thus storing magnetic energy, until a kink instability
develops \citep{2013ApJ...771...76R}. Similarly, an isolated $z$-boundary vortex,
even in the presence of strong nonlinear dynamics in the corona, can store energy
at large spatial scales via an inverse cascade of energy, with subsequent
energy release events in the micro-flare range \citep{2013ApJ...771...76R}.
Similarly also $z$-boundary shear motions, isolated or stronger than
surrounding motions, can store a large amount of energy as sheared
magnetic field lines that  can subsequently be released impulsively
\citep{2005ApJ...622.1191D, 2009ApJ...704.1059D, 2010ApJ...722...65R}.

We want to stress the fact that the amount of energy entering from the footpoints is an \emph{outcome} of
the simulation since such energy depends on $B_{perp}$ that cannot be specified as a
boundary condition. That means that the loop \emph{nonlinear} dynamics
determines how much energy can be injected into the system and that the
``heating function" cannot be assigned a priori. Furthermore the almost perfect correlation
between   $T_{max}$ and $J_{max}$, confirming that the peaks in temperature are due to the
local enhancement of the current, shows that the heating is due to local phenomena. These
phenomena are the results of the complex perpendicular dynamics driven by $z$-boundary
motions which induce a local increase of the heating which in this framework is due
exclusively to magnetic reconnection. Most of the dissipation occurs within localized current
sheets which disrupt rapidly on Alfv\'en time-scales when their perpendicular size decreases
to the smallest spatial scale present in our simulation. It is interesting to notice the good
correlation between the behavior of the Poynting Flux and the  energy dissipation. The two
curves are very similar and shifted in time which means that when the system admits a bigger
average energy flux from the two bases, current starts piling up locally leading to an increase
of total energy dissipated and to a formation and disruption of localized current sheets. The
average Poynting flux depends on the length of the loop.
The time averaged flux is of the order of $1\times10^4$\,J~m$^{-2}$~s$^{-1}$
for the hotter loop (case C) and $5\times10^2$\,J~m$^{-2}$~s$^{-1}$  for the
cooler one (case A ).
The Poynting flux thus increases almost quadratically with magnetic guide field intensity
as in previous reduced MHD studies \citep{2008ApJ...677.1348R}.

As already mentioned the resolution of our simulations is coarse compared with the real
scales present in the corona and consequently we are using values of resistivity and viscosity
which are much higher than the real ones, which are unachievable using present computers. 
We have verified that doubling the numerical
resolution, and therefore halving the resistivity and viscosity, changes the
significant results only weakly (as expected ).
Previous reduced MHD simulations have shown that total dissipation
rates and Poynting flux increase with increasing resolutions, saturating at resolutions of about
$256^2 \times 128$.  In a fully developed turbulent regime dissipation rates are not expected
to depend on the Reynolds numbers beyond a certain threshold.
Previous simulations suggest that the resolutions adopted in this paper
are below but not too far from such a threshold
\citep{2008ApJ...677.1348R, 2010ApJ...722...65R}, confirming
that the results presented here for the radiative losses are realistic.
The challenging investigations of the dynamics in the high
Reynolds regime and their impact on observations - expected to increase intermittency effects, and ultimately require kinetic calculations on the dissipation scale, is left to dedicated future works.
Most phenomenological studies of coronal heating have so far concentrated on the thermodynamic response
of the coronal plasma
using one-dimensional hydrodynamic loop models with a prescribed heating function.
These models have been in common use in coronal physics for
over thirty years, and have been fundamental in providing a basic
framework for a wide variety of coronal
phenomena, including loop temperature distributions
\citep{1978ApJ...220..643R}, prominence formation \citep{1982ApJ...254..349O}
and, more germane to this paper, coronal heating
\citep{2014LRSP...11....4R,2006SoPh..234...41K,2004ApJ...605..911C}.

In the one-dimensional model heating is often represented as a constant
\citep[see, e.g.,][]{1981A&A....97...27C, 1982A&A...105L...1T}.
It also can be generalized to be a function of some combination of mass density,
temperature and thermal pressure. In the latter cases the heating can be both
spatially and temporally dependent.
A particular case of interest is that of impulsive heating, in which the heating function is turned
on for some span of time, and then shut off to allow the loop to evolve to a new equilibrium.
The heating can be localized in the photosphere or appear as bursts in the corona.

The main limitation of one-dimensional models
is that whatever functional dependence is chosen, the heating remains an
{\it ad hoc} function and the main task of the researcher is to see which of these
dependencies provides the best fit to observations.  The chosen functional dependence is
thought to lead to some understanding of which mechanism heats the corona,
but the link with coronal heating theories and models remains
essentially undetermined.
For instance in the case of the Parker model investigated in this paper it is not
obvious at all what heating function in 1D models would represent it better.
One might be tempted, for example, to use a heating function varying very slowly with time,
since photospheric motions have a very low frequency compared to the
fast Alfv\'en crossing time. But previous reduced MHD and our simulations
show that the system develops turbulent dynamics, with the timescale
of the system strongly decreasing at smaller scales
\citep{2007ApJ...657L..47R, 2008ApJ...677.1348R, 2011PhRvE..83f5401R}.
This is independently confirmed by the recent finding that in such
systems current sheets form on very fast ideal timescales, with
their thickness thinning at least exponentially and reaching the
dissipative scale in about an Alfv\'en crossing time \citep{2013ApJ...773L...2R}.
Such thinning current sheets have also been shown to be
unstable to tearing modes with ideal growth rates \citep{2014ApJ...780L..19P},
with the formation of many magnetic islands and X-points
and the complex dynamics of so-called super-tearing  or plasmoid instability
\citep{bss78, bisk86, 2007PhPl...14j0703L, 2008PhRvL.100w5001L} ensuing.
Determining the equivalent heating function
for 1D simulations from this framework of coronal heating is therefore
a complex task. In particular such heating function for the Parker model has never been investigated, and therefore
the 1D hydrodynamic models have not been \emph{de facto}
able to test it \citep{2015TESS....120308K}.

As observational evidence has accumulated that many loops are not isothermal, it has
become apparent that coronal loops cannot be modeled using a single flux tube
\citep{2010ApJ...723.1180S}.
The narrow temperature distributions \citep{2008ApJ...686L.131W} and their transient
nature \citep{2009ApJ...695..642U} point to multiple structures and coherence.
In an effort to account for these observations, refined multi-strand models
\citep[e.g][]{2009ASPC..415..221K}  have been developed, in which an
ensemble of one dimensional loops is assembled in an attempt to construct a
three-dimensional loop. Our numerical simulations give strong support
to a multi-temperature coronal loop structure whose specific temperature
distribution is likely to depend on the loop parameters, similar
to the Emission Measure Distribution shown in Figure~10,
that we plan to further investigate in future work.

It is important to emphasize that,
as far as the thermodynamics is concerned, we are solving the same equations that are used in a  reduced form in one-dimensional models. Looking at the big differences in
temperature appearing in the 2D plots in the mid-plane, it is easy to understand that the
temperature profiles along different field lines originating in different points of the mid-plane
can differ in a very substantial way since for all field lines the temperature at the footpoints is
$10^4$ K. No field line can be considered representative of what happens in the loop.
The limitation to one spatial dimension would leave out the self-consistent nonlinear
dynamics with the most significant energy transfers, responsible for the formation
of current sheets and thus the energy deposition at small scales, occurring in the
perpendicular directions.  Additionally for energy to be transferred from the magnetic
field to the plasma magnetic reconnection must occur,
hence magnetic field lines are constantly being broken and reconnected
\citep{2007ApJ...662L.119S} strongly impacting the energy
distribution along different strands.

\acknowledgments
We thank an anonymous referee for helpful remarks.
We thank J. P. Dahlburg and J. M. Laming for helpful conversations.
This research was supported in part by NRL 6.2 funds, the NASA SR\&T program,
and by NASA through subcontracts with the Jet Propulsion Laboratory, California
Institute of Technology.
Computational resources supporting this
work were provided by LCP\&FD, and in part by the DOD HPCMP.
AIA and HMI data are courtesy of NASA/SDO and the AIA  and HMI science teams.
Hinode is a Japanese mission developed and launched by ISAS/JAXA,
with NAOJ as domestic partner and NASA and STFC (UK) as international partners. It is
operated by these agencies in co-operation with ESA and NSC (Norway).

\end{document}